\documentclass[11pt]{article}
\usepackage{latexsym}
\usepackage{amsmath,amssymb,amsthm,hyperref}
\usepackage[english]{babel}
\usepackage[dvips]{graphicx}
\usepackage{hyperref}

\def\AFOUR{%
\setlength{\textheight}{8.5in}%
\setlength{\textwidth}{5.75in}%
\setlength{\topmargin}{-0.375in}%
\hoffset=-.5in%
\renewcommand{\baselinestretch}{1.17}%
\setlength{\parskip}{6pt plus 2pt}%
}

\AFOUR                                           %%%   ***

\expandafter\ifx\csname amssym.def\endcsname\relax \else\endinput\fi
\expandafter\edef\csname amssym.def\endcsname{%
       \catcode`\noexpand\@=\the\catcode`\@\space}
\catcode`\@=11
\def\undefine#1{\let#1\undefined}
\def\newsymbol#1#2#3#4#5{\let\next@\relax
 \ifnum#2=\@ne\let\next@\msafam@\else
 \ifnum#2=\tw@\let\next@\msbfam@\fi\fi
 \mathchardef#1="#3\next@#4#5}
\def\mathhexbox@#1#2#3{\relax
 \ifmmode\mathpalette{}{\m@th\mathchar"#1#2#3}%
 \else\leavevmode\hbox{$\m@th\mathchar"#1#2#3$}\fi}
\def\hexnumber@#1{\ifcase#1 0\or 1\or 2\or 3\or 4\or 5\or 6\or 7\or 8\or
 9\or A\or B\or C\or D\or E\or F\fi}

\font\tenmsa=msam10
\font\sevenmsa=msam7
\font\fivemsa=msam5
\newfam\msafam
\textfont\msafam=\tenmsa
\scriptfont\msafam=\sevenmsa
\scriptscriptfont\msafam=\fivemsa
\edef\msafam@{\hexnumber@\msafam}
\mathchardef\dabar@"0\msafam@39
\def\dashrightarrow{\mathrel{\dabar@\dabar@\mathchar"0\msafam@4B}}
\def\dashleftarrow{\mathrel{\mathchar"0\msafam@4C\dabar@\dabar@}}

\def\ulcorner{\delimiter"4\msafam@70\msafam@70 }
\def\urcorner{\delimiter"5\msafam@71\msafam@71 }
\def\llcorner{\delimiter"4\msafam@78\msafam@78 }
\def\lrcorner{\delimiter"5\msafam@79\msafam@79 }
\def\yen{{\mathhexbox@\msafam@55}}
\def\checkmark{{\mathhexbox@\msafam@58}}
\def\circledR{{\mathhexbox@\msafam@72}}
\def\maltese{{\mathhexbox@\msafam@7A}}
\def\circledS{{\mathhexbox@\msafam@73}}

\font\tenmsb=msbm10
\font\sevenmsb=msbm7
\font\fivemsb=msbm5
\newfam\msbfam
\textfont\msbfam=\tenmsb
\scriptfont\msbfam=\sevenmsb
\scriptscriptfont\msbfam=\fivemsb
\edef\msbfam@{\hexnumber@\msbfam}
\def\Bbb#1{{\fam\msbfam\relax#1}}
\def\widehat#1{\setbox\z@\hbox{$\m@th#1$}%
 \ifdim\wd\z@>\tw@ em\mathaccent"0\msbfam@5B{#1}%
 \else\mathaccent"0362{#1}\fi}
\def\widetilde#1{\setbox\z@\hbox{$\m@th#1$}%
 \ifdim\wd\z@>\tw@ em\mathaccent"0\msbfam@5D{#1}%
 \else\mathaccent"0365{#1}\fi}
\font\teneufm=eufm10
\font\seveneufm=eufm7
\font\fiveeufm=eufm5
\newfam\eufmfam
\textfont\eufmfam=\teneufm
\scriptfont\eufmfam=\seveneufm
\scriptscriptfont\eufmfam=\fiveeufm
\def\frak#1{{\fam\eufmfam\relax#1}}

\csname amssym.def\endcsname

\makeatletter
\def\section{\@startsection {section}{1}{\z@}{-3.5ex plus -1ex minus
 -.2ex}{2.3ex plus .2ex}{\large\sc}}
\def\subsection{\@startsection{subsection}{2}{\z@}{-3.25ex plus -1ex minus
 -.2ex}{1.5ex plus .2ex}{\normalsize\sc}}
\makeatother

\makeatletter
\@addtoreset{equation}{section}

\makeatother

\newcommand{\nc}{\newcommand}
\newcommand{\rnc}{\renewcommand}

\nc{\subs}[1]{{\vspace*{0.5cm}}%
{\noindent\underline{\small\sc #1}}{\addcontentsline{toc}{subsubsection}{#1}}%
{\vspace*{0.3cm}}}

\nc{\subss}[1]{{\vspace*{0.5cm}}%
{\noindent\underline{\small\sc #1}}%
{\vspace*{0.3cm}}}

\nc{\chap}[1]{{\clearpage}%
\begin{center}%
{\noindent\underline{\large\sc #1}}{\addcontentsline{toc}{section}{#1}}%
\end{center}%
{\vspace*{0.3cm}}}

\nc{\be}{\begin{equation}}
\nc{\ee}{\end{equation}}
\nc{\bea}{\begin{eqnarray}}
\nc{\eea}{\end{eqnarray}}

\nc{\trac}[2]{{\textstyle\frac{#1}{#2}}}

\nc{\ex}[1]{\mbox{e}^{\,\textstyle#1}}

\nc{\CC}{\Bbb{C}}
\nc{\HH}{\Bbb{H}}
\nc{\PP}{\Bbb{P}}
\nc{\ZZ}{\Bbb{Z}}
\nc{\II}{\Bbb{I}}
\nc{\EE}{\Bbb{E}}

\rnc{\a}{\alpha}
\rnc{\b}{\beta}
\rnc{\d}{\delta}
\nc{\ga}{\gamma}
\nc{\la}{\lambda}
\nc{\f}{\phi}
\nc{\p}{\psi}
\rnc{\c}{\chi}
\nc{\om}{\omega}
\nc{\Om}{\Omega}

\nc{\symx}{\circledS}
\newsymbol\smallsmile 1360
\newsymbol\smallfrown 1361
\nc{\ad}{\mathop{\mbox{ad}}\nolimits}
\nc{\tr}{\mathop{\mbox{tr}}\nolimits}
\nc{\Tr}{\mathop{\mbox{Tr}}\nolimits}
\nc{\Det}{\mathop{\mbox{Det}}\nolimits}
\rnc{\det}{\mathop{\mbox{det}}\nolimits}
\nc{\rk}{\mathop{\mbox{rk}}\nolimits}
\nc{\del}{\partial}
\nc{\diag}{\mathop{\mbox{diag}}\nolimits}
\nc{\ra}{\rightarrow}
\nc{\Ra}{\Rightarrow}
\nc{\LRa}{\Leftrightarrow}
\nc{\lra}{\leftrightarrow}
\nc{\ot}{\otimes}
\rnc{\ss}{\subset}
\nc{\nul}{\noindent\underline}
\nc{\non}{\nonumber\\}
\nc{\mat}[4]{\left(\begin{array}{cc}#1&#2\\#3&#4\end{array}\right)}
\rnc{\lg}{\frak{g}}
\nc{\nam}{\nabla_{\mu}}
\nc{\nan}{\nabla_{\nu}}
\nc{\dx}{\dot{x}}
\nc{\dxl}{\dot{x}^{\la}}
\nc{\dxm}{\dot{x}^{\mu}}
\nc{\dxn}{\dot{x}^{\nu}}
\nc{\ddx}{\ddot{x}}
\nc{\ddxm}{\ddot{x}^{\mu}}
\nc{\ddxn}{\ddot{x}^{\nu}}
\nc{\dxi}{\dot{\xi}}
\nc{\ddxi}{\ddot{\xi}}

\def\la{\label}
\def \p {\phi}

\theoremstyle{plain}
\newtheorem{teo}{Theorem}[section]

\theoremstyle{definition}

\newcommand{\der}{\partial}
\newcommand{\e}{\mathrm{e}}
\newcommand{\ii}{\mathrm{i}}
\newcommand{\dd}{\mathrm{d}}
\newcommand{\vb}[1]{\mathbf{#1}}

\newcommand{\eps}{\varepsilon}

\newcommand{\RR}{\mathbb R}

\newcommand{\tend}{\rightarrow}
\newcommand{\then}{\Rightarrow}

\begin{document}

\rightline{SISSA/49/2005/EP}
%\rightline{Version of \today}

\vfill

\begin{center}
{\Large\sc Singularities and closed time-like curves\\ in type IIB
1/2 BPS geometries}
\end{center}
\vspace{0.3cm}

\begin{center}
{\large Giuseppe Milanesi and Martin O'Loughlin}
\end{center}

\vskip 0.05 cm \centerline{\it S.I.S.S.A. Scuola Internazionale
Superiore di Studi Avanzati} \centerline{\it Via Beirut 4, 34014
Trieste, Italy} \centerline{INFN, Sezione di Trieste}

\vskip -2.0 cm

\begin{center}
{\bf Abstract}
\end{center}
\vskip -0.1 cm We study in  detail the moduli space of solutions
discovered in LLM relaxing the constraint that guarantees the
absence of singularities. The solutions fall into three classes,
non-singular, null-singular and time machines with a time-like
naked singularity. We study the general features of these metrics
and prove that there are actually just two generic classes of
space-times - those with null singularities are in the same class
as the non-singular metrics. $AdS/CFT$ seems to provide a dual
description only for the first of these two types of space-time in
terms of a unitary $CFT$ indicating the possible existence of a
chronology protection mechanism for this class of geometries.
\vfill

\newpage
\begin{small}
\tableofcontents
\end{small}

%\newpage
\setcounter{footnote}{0}

\section{Introduction}

In \cite{bubbling} a class of type IIB 1/2  BPS solutions has been
constructed together with their $CFT$ duals. This construction has
inspired interesting work in various directions,
\cite{Liu:2004hy,Liu:2004ru,Martelli:2004xq,Chong:2004ce,Ghodsi:2005ks,Ebrahim:2005uz,Sheikh-Jabbari:2005mf,Mukhi:2005cv,Spalinski:2005ha,Bena:2005va}
and \cite{Mandal:2005wv,Grant:2005qc,Balasubramanian:2004nb}. The
basic trick of \cite{bubbling} is to note that assuming a certain
amount of symmetry in the ansatz for metric and five-form field
strength, and demanding that the geometry has a Killing spinor,
the remaining equations of motions reduce to an elliptic equation
for a scalar function $z$ on $\RR^2\times\RR^+$. The value of
$\rho = 1/2 - z$ on the 2-plane boundary of $\RR^2\times\RR^+$ can
be interpreted as a semiclassical fermion density, thus providing
 direct contact to the $CFT$ dual Yang-Mills theory on $\RR \times
S^3$\cite{Berenstein:2004kk,Corley:2001zk}. Indeed if this density
takes on only the values 0 and 1 then the solutions are guaranteed
to be singularity free space-times.

In this paper we consider the most general allowed (on the
supergravity side) boundary conditions for the elliptic equation.
This means that we study the full set of moduli of this sector of
supergravity that consists of metrics asymptotic to $AdS_5\times
S^5$, with an $SO(4)\times SO(4)$ isometry group and preserving
half of the supersymmetry of type IIB string theory. The
supergravity solutions in general will be singular. The spacetime
singularities appearing are always naked and fall into two
distinct classes: null and timelike. The null ones can be
considered as seeded by a fermion density between 0 and 1 and are
already considered in the literature, see for example
\cite{Myers:2001aq,Buchel:2004mc,Horava:2005pv,Gubser:2004xx,Bak:2005ef}
together with the possible \emph{local} quantum effects
responsible for their resolution - the singularity is due to an
average over configurations of $N$ fermions in a gas with average
density less than one. An individual configuration with the same
asymptotics can actually be seen to have as source a collection of
$N$ giant gravitons \cite{McGreevy:2000cw,Hashimoto:2000zp}
separated one from the other. In the supergravity theory, the
resolution of the singularity thus appears as a sort of space-time
foam \cite{Balasubramanian:2005kk} while in the dual $CFT$ one
sees that such a configuration corresponds to the Coulomb branch
of the theory.

The $AdS/CFT$ correspondence has maybe something more interesting
to tell us about the fate of the timelike singularities. The
solutions with this kind of singularity are highly
``pathological'': they have closed timelike curves passing through
\emph{any point} of the spacetime and they include
unbounded from below negative mass excitations of $AdS_5\times
S^5$.

It has already been conjectured,
\cite{Gubser:2000nd,Myers:1999ps,Horowitz:1995ta}, that geometries
with these features should be considered as truly unphysical via
\emph{global} considerations in the setting of a full quantum
theory of gravity. The $AdS/CFT$ correspondence applied to the
space-times of \cite{bubbling} suggests one particular mechanism
for the global removal of solutions containing timelike
singularities. The deformations of the geometry which produce
these singularities apparently correspond to negative dimension
operators in the dual field theory. The unitarity of the
representations of the superalgebra $SU(2,2|4)$ \cite{dobrev}
indicate in particular that unitary operators must have a positive
conformal dimension. Our observations indicate that there should
actually exist a general proof of the chronology protection
conjecture \cite{Hawking:1991nk} in this sector of supergravity .
A first indication of this mechanism linking unitarity to
chronology protection can be found in \cite{Breckenridge:1996is}
and in the current context in
\cite{Caldarelli:2004mz,Boni:2005sf}.

Extending these works, in this paper we prove that closed timelike
curves (CTCs) are unavoidable in any solution with a timelike
singularity and that they are excluded in the case of regular and
null singular solutions, these being the spacetimes that can be
represented in terms of dual fermions, a result anticipated but
not proven in \cite{Caldarelli:2004mz}. This provides a clear
division between these two classes of singular spacetimes which is
also reflected in the two different mechanisms responsible for the
resolution of their respective spacetime singularities.

In Section 2 we review the construction of \cite{bubbling} and we
show the most general allowed boundary conditions for a
supergravity solution satisfying the symmetry requirements.
We clarify the role of the boundary conditions in
determining the radius of the asymptotic $AdS_5\times S^5$ and we
show the relation between the boundary conditions and the
appearance of spacetime singularities.

In Section 3 we exhibit some examples of singular supergravity
solutions and we uncover some of their properties such as CTCs
and peculiar geometric features. In particular
we exhibit unbounded from below (for fixed $AdS$ radius)
negative mass excitations of $AdS_5\times S^5$.

In Section 4 we show that most of the interesting features of the
examples in Section 3, regarding mainly the appearance and the
properties of CTCs, are generic for the case of solutions with
timelike singularities. Moreover we prove a theorem which clearly
relates the appearance of CTCs to the boundary conditions responsible
for timelike singularities.

In section 5 we return to a discussion of the meaning of these
results, and in particular the possibility of proving the
chronology protection conjecture for this class of geometries, by
showing that the $AdS/CFT$ correspondence relates naked time
machines to non-unitarity in the $CFT$.

In the Appendix we show that there is just one plane wave
geometry, the maximally supersymmetric one of \cite{Blau:2001ne},
that can be obtained from the construction presented in
\cite{bubbling}.

\section{LLM construction}\label{LLM constr and
generalization} In the first part of this section we review the
construction of \cite{bubbling} in a language adapted to the
considerations that follow in the rest of this paper.

In \cite{bubbling} a class of BPS solutions of
type IIB supergravity is constructed. This is the most general
class of BPS solutions in type IIB supergravity with $SO(4)\times
SO(4)$ isometry, one timelike
Killing vector and a non-trivial self-dual 5-form field strength $F_{(5)}$.
The solutions are given by
\begin{gather}
  \dd s^2
  =-h^{-2}(\dd t + V_i \dd x^i)^2+h^2(\dd y^2 + \delta_{ij}\dd x^i\dd x^j)+
y\e^G\dd\Omega_3^2+y\e^{-G}\dd\widetilde{\Omega}_3^2
  \label{geometry}\\
  F_{(5)}=F_{\mu\nu}\dd x^\mu\wedge\dd x^\nu\wedge\dd\Omega+
\widetilde{F}_{\mu\nu}\dd x^\mu\wedge\dd
  x^\nu\wedge\dd\widetilde{\Omega}\\
   F=\e^{3G}\ast_4\widetilde{F}\\
\end{gather}
with $y\geq0$.\\
We can define a function $z=z(x_1,x_2,y)$ which determines the
entire solution (up to choice of gauge that we discuss below),
\begin{gather}
  z\equiv \frac12\tanh G\\\label{h-2}
  h^{-2}=2y\cosh G=\frac y{\sqrt{(1/2-z)(1/2+z)}}\\
  \dd V=\frac1y\ast_3\dd z\label{dv=1/y dz}\\
  F= \dd (B_t(\dd t + V)) +\dd \hat B\\
    \tilde{F}= \dd (\tilde{B}_t(\dd t + V)) +\dd \hat
    {\tilde{B}}\\
    B_t=-\frac14 y^2 \e^{2G}\qquad \tilde B_t=-\frac14 y^2
    \e^{-2G}\\
    \dd \hat B= -\frac14 y^3 \ast_3\dd\left(\frac{1/2+z}{y^2}\right)
\qquad     \dd \hat{\tilde B}= \frac14 y^3
    \ast_3\dd\left(\frac{1/2-z}{y^2}\right)
\end{gather}
where $\ast_n$ indicate the Hodge dual in $n$ flat dimensions. \\
For the consistency of \eqref{dv=1/y dz} we must have
\begin{equation}\label{laplace for z}
  (\der_1^2+\der_2^2)z+y\der_y(\frac1y\der_y z)=0
\end{equation}
The solutions for $z$ are determined by boundary conditions in the
$\{x_1,x_2,y\}$ space as we will now discuss.

\subsection{Boundary conditions}\label{Boundary conditions}

The solution is well defined for $z$ restricted to the range
\begin{equation}
-1/2\leq z\leq1/2
\end{equation}
Equation \eqref{laplace for z} implies that $z$ takes its maximum
and minimum on the boundary of its domain of
definition\footnote{The equation \eqref{laplace for z} can be
rewritten as
\begin{equation}
\left(  \der_1^2+\der_2^2+\der_y^2 - \frac 1 y \der_y \right) z =
0
\end{equation}
Assume that $Q$ is an internal stationary point of $z$, then
clearly $\der_y z (Q) = 0$. The equation for $z$ implies that
$\left( \der_1^2+\der_2^2+\der_y^2\right) z (Q) = 0$, and thus $Q$
cannot be a maximum (nor  a minimum).}
 $\Sigma \subset \RR^2\times \RR^+$. A
solution of the supergravity equations is thus specified by a
choice of $\Sigma$, and by a function $z_0$ defined on $\der
\Sigma$ such that
\begin{gather}
  z=z_0\qquad\text{on}\,\,\der \Sigma\\
  -1/2\leq z_0\leq 1/2\notag
\end{gather}
Following \cite{bubbling} one can easily show that if $\Sigma$
extends to infinity and $z$ goes to either $1/2$ or $-1/2$ for
$r^2=x_1^2+x_2^2+y^2\tend\infty$, the solution is asymptotically
$AdS_5\times S^5$. Changing $z$ into $-z$ is a symmetry of the
solution and thus we assume for definiteness
\begin{equation}
  z\tend \frac12\,\,\text{for}\,\,r^2=x_1^2+x_2^2+y^2\tend\infty
\end{equation}

We call $\der \Sigma_0$ the intersection of $\der \Sigma$ with the
$y=0$ plane, and $\der \hat \Sigma=\der \Sigma \setminus \der
\Sigma_0$. We note that if $z_0\neq\pm\frac12$ on $\der \hat
\Sigma$ then the metric can be analytically continued as far as
$y=0$ or $z=\pm \frac12$. In general, after analytically
continuing the solution, we have a larger ``maximal'' domain
$\Sigma' \supset \Sigma$ where $- \frac 1 2 \leq z\leq \frac 1 2$.

For convenience we will call again $\Sigma$ this \emph{maximal}
domain of definition. The most general asymptotically $AdS_5\times
S^5$ solution of the supergravity equations is then specified by
the domain ${\Sigma}$ and a function $z_0$ on $\der \Sigma$

\begin{equation}\label{boundary conditions for z}
\begin{cases}
  -\frac12\leq z_0\leq\frac12 \quad &\text{on}\,\,\der\Sigma_0\\
    z_0=\pm \frac12 \quad &\text{on}\,\,\der\hat\Sigma\\
    z_0\tend\frac12 \quad &\text{for}\,\,r\tend\infty
  \end{cases}
\end{equation}
as illustrated in Figure 1.

\begin{figure}
\begin{center}
\includegraphics[width=7 cm]{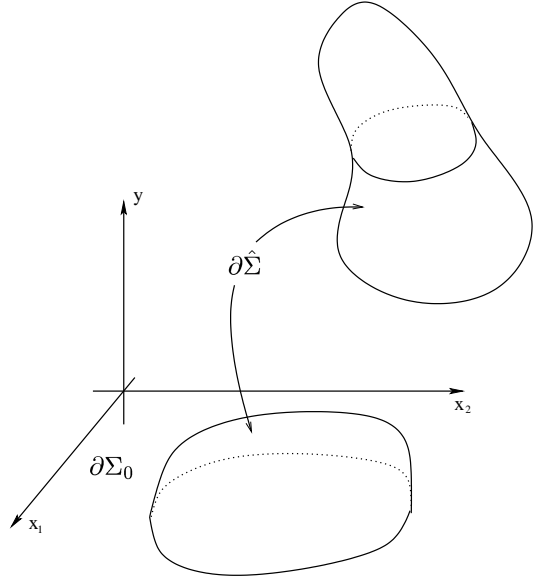}
\end{center}
\caption{The domain of definition $\Sigma$}
\end{figure}

We define a new function $\Phi$
\begin{equation}
  \Phi\equiv \frac {\frac12-z}{y^2}
\end{equation}
The equation for $z$ is equivalent to the Laplace equation for
$\Phi$ on a flat six dimensional space of the form
$\RR^2\times\RR^4$ where $x_1,x_2$ are the coordinates on the
$\RR^2$ and $y$ is the radius for spherical coordinate on the
$\RR^4$. Since \eqref{laplace for z} and the definition of $\Phi$
are singular for $y=0$, Dirichlet boundary conditions for $z$ on
$y=0$ take the role of charge sources for $\Phi$ located at $y=0$.
Thus $\Phi$ satisfies the equation
\begin{equation}\left\{\begin{split}
  \label{laplace for phi}
  &(\der_1^2+\der_2^2)\Phi+\frac 1{y^3}\der_y(y^3\der_y
  \Phi)=\ast_6\dd\ast_6\dd\Phi=-4\pi^2
(\frac12-z_0)\delta^{(4)}(y)\chi(\Sigma_0)\\
  &\Phi=\frac{\frac12-z_0}{y^2} \,\,\text{on}\,\der \hat
  \Sigma\end{split}\right.
\end{equation}
where
\begin{equation}
  \chi(\Sigma_0)(x_1,x_2)=\begin{cases}
    1 &\text{if}\,\,(x_1,x_2)\in \Sigma_0\\
    0 &\text{otherwise}
  \end{cases}.
\end{equation}

The forms $B,\tilde B$ and $V$ are defined up to a gauge
transformation. From now on we will use the following convenient
gauge for $V$
\begin{equation}
  \dd \ast_3 V = 0 \iff \der_1V_1+\der_2V_2=0
\end{equation}

\subsection{Asymptotic behaviour}\label{Asymptotic behaviour}
The boundary conditions \eqref{boundary conditions for z} imply that
\begin{equation}
  \Phi\tend\frac A {(x_1^2+x_2^2+y^2)^2}=\frac {A}{r^4},\qquad r\tend\infty,
\qquad A>0.
\end{equation}
Integrating \eqref{laplace for phi} we obtain
\begin{equation}
  A \, 4\pi^3=\int_{S^5(r)}(-\ast_6\dd \Phi) = \int_{\hat\Sigma^5}
(-\ast_6\dd\Phi) + 4\pi^2 \int_{\der \Sigma_0}
  \left(\frac12-z_0(x_1,x_2)\right)\dd x_1 \dd x_2
\end{equation}
where $S^5(r)$ is a 5-sphere of radius $r$ centered on the origin,
and $\hat\Sigma^5$ is the 5-manifold obtained by the fibration in
spherical coordinates of a 3-sphere $S^3(y)$ over $\der \hat
\Sigma$.\\
Going to polar coordinates $R,\varphi$ in the $\{x_1,x_2\}$
sections and for $R^2+y^2\tend\infty$, $V$ has the asymptotic
behaviour
\begin{equation}
  V\approx V_\varphi\dd\varphi\qquad V_\varphi\approx - A \frac{R^2}
{(R^2+y^2)^2}.
\end{equation}
In \cite{bubbling} it has been shown that the quantity $A$
determines the radius of the asymptotic $AdS_5\times S^5$
\begin{equation}
  R^2_{AdS_5}=R^2_{S^5}=A^{1/2}
\end{equation}

In the asymptotic region we can construct a smooth five
dimensional manifold $\tilde\Lambda_5$ by fibering the three
sphere $\tilde S^3$ over a surface $\tilde \Lambda_2$.\\
The topology of $\tilde \Lambda_5$ is asymptotically $S^5$. The
flux of the five form through this surface is given by
\begin{equation}\label{generic flux of F5}
  N=-\frac1{4\pi^4l_P^4}\int_{\tilde\Lambda_5}\dd \hat{
  \tilde B}\wedge\dd\tilde\Omega_3=-\frac
  1{16\pi^4l_P^4}\int_{\tilde\Lambda_5}\ast_6\dd \Phi=\frac 1{4\pi l_P^4}
  A=\frac 1{4\pi l_P^4}R_{AdS_5}^4
\end{equation}
which agrees with the standard formula for the relation between
the radius of $AdS_5$ and the flux of $F_{(5)}$.

The mass of the excitation of $AdS_5\times S^5$ can be computed by
looking at subleading terms in the expansion of $\Phi$ around
$r\tend\infty$.

\subsection{Regular solutions and dual picture}
If we choose $\Sigma=\RR^2\times\RR^+$ the solution can be written
as\footnote{Note that\begin{equation*}
\lim_{y\tend0}\frac{y^2}\pi\frac{1}{(x_1^2+x_2^2+y^2)^2}=\delta^{(2)}(x_1,x_2)
\end{equation*}}
\begin{gather}\label{solution for Phi,V for generic z0}
  z=\frac12-\Phi \,
  y^2=\frac 12 - \frac{y^2}\pi\int\frac{\rho(x_1',x_2')\dd^2x'}
{[(x_1-x_1')^2+(x_2-x_2')^2+y^2]^2}\\
  V_i=-\frac1\pi\epsilon_{ij}\int\frac{(x_j-x_j')\rho(x_1',x_2')\dd^2
  x'}{[(x_1-x_1')^2+(x_2-x_2')^2+y^2]^2}
\end{gather}
with
\begin{equation}
  \rho (x_1,x_2) = \frac 12  - z_0 (x_1,x_2)
\end{equation}
 According to \cite{bubbling}, in the dual field theory
these excitation of $AdS_5\times S^5$
 are described by $N$ free fermions. The plane $y=0$ can be
identified with the phase space of the dual fermions and the
function $\rho(x_1,x_2)$ can be identified with the semiclassical
density of these fermions.

It can be shown that the metric is regular if
$\Sigma=\RR^2\times\RR^+$ and $z_0$ takes the values $\pm 1/2$ on
the $y=0$ plane \cite{bubbling}. In these cases $\rho$ is non
vanishing just inside the ``droplets" where $z_0=-1/2$
\begin{equation}
  \rho=\begin{cases}
     \beta = 1 \,&\text{inside the droplets}\\
0 \, &\text{outside}
  \end{cases}
\end{equation}
Since we have assumed that $z\tend 1/2$ at infinity, we can always
find a circle large enough to encircle all ``droplets''. With
these boundary conditions $z$ is given by
\begin{equation}
  z= \frac12-\frac{y^2}\pi\int_{\mathcal
  D}\frac{\dd^2x'}{[(x_1-x_1')^2+(x_2-x_2')^2+y^2]^2}
\end{equation}
$\mathcal D$ being the union of the droplets where $z=-1/2$. The
$V$ form can be written
\begin{equation}
V_i=-\frac1\pi\epsilon_{ij}\int_{\mathcal
  D}\frac{(x_j-x_j')\dd^2
  x'}{[(x_1-x_1')^2+(x_2-x_2')^2+y^2]^2}
  \end{equation}
  The determinant of the sections $\{x_1,x_2,y\}$ is given by
  \begin{equation}
    \tilde g = h^4-V_1^2-V_2^2=\frac {1/4 - z^2} {y^2} - V^2
  \end{equation}
Note that here and in the following $V^2$ is formed by contracting
indices using the Kronecker delta, i.e. $V^2\equiv V_1^2+V^2_2$.
Theorem \ref{necessary and sufficient condition for CTCs} of
Section 4 states that for any\footnote{Even extended to infinity,
that is relaxing the hypotheses $z\tend 1/2$ for $r\tend\infty$
and allowing more general asymptotics than $AdS_5\times S^5$}
$\mathcal D$, $\tilde g\geq 0$ and the $\{x_1,x_2,y\}$ sections do
not contain time-like directions. This guarantees in particular
that the original LLM solutions are free of CTCs and are ``good''
supergravity solutions.

 From the analysis of Section \ref{Asymptotic behaviour}, we can
deduce that the radius of the asymptotic $AdS_5\times S^5$ is
given by
\begin{equation}
  R_{AdS}^4=\frac{\mathcal S}\pi=A
\end{equation}
where
\begin{equation}
  \mathcal S = \int_{\mathcal D}\dd x_1 \dd x_2
\end{equation}
is the total area of all droplets where $z=-\frac12$ ($\rho=1$). The
quantization of the flux \eqref{generic flux of F5} gives the
quantization condition on the area of the droplets
\begin{equation}
  \mathcal S = 4\pi^2 l_p^4 N
   \end{equation}
If $\mathcal D$ consists of one single circular droplet then
the spacetime is precisely $AdS_5\times S^5$. For a generic
set of droplets $\mathcal D$ the mass (and the angular momentum)
of the excitation is given by
  \begin{equation}
    M = J = \frac 1 {8\pi^2 l_P^8}\left[\frac 1
{2\pi}\int_{\mathcal D} (x_1^2+x_2^2)\dd^2
    x-\left(\frac 1 {2\pi} \int_{\mathcal D} \dd^2 x\right)^2\right]\geq0
\end{equation}
The origin of the coordinates is chosen such that the dipole vanishes,
that is,
\begin{equation}
\int_{\mathcal D} x_i \dd^2 x = 0.
\end{equation}
Not surprisingly one can
show by a direct calculation that the equality ($M=J=0$) holds for
 a single disk. Given any $\mathcal D$ we can build a disk
 $C_{\mathcal D}$
 of the same area. The first term is clearly larger for $\mathcal
 D$ than for $C_{\mathcal D}$ and thus in general $M>0$ for the non-singular
solutions.

\subsection{More general boundary conditions and singularities}

In all cases with boundary conditions different from the ones
studied in \cite{bubbling} we have spacetime singularities.

It is easy to see that the solutions have a naked time-like
singularity when  $\der \hat \Sigma $ is non-empty. Consider a
surface in the region $y>0$ on which $z= -1/2$ (the discussion
does not change in any substantial way if instead we took
$z=1/2$). Choose a point $Q$ on this surface and define a
coordinate $\epsilon$ in the $\{x_1,x_2,y\}$ space orthogonal to
this surface such that $z = -1/2 + \alpha\epsilon$ for some
positive constant $\alpha$. Complete $\epsilon$ to a new
orthogonal coordinate system by introducing two coordinates $v_i$
with origin at $Q$. This is just an orthogonal transformation and
translation of the original coordinate system. At $Q$ we can
assume that $V$ is finite with a power series expansion away from
this point. The subleading terms in this expansion are not
important for studying the singularity. We also define a new time
coordinate near Q by $T= t+V_i(Q)x_i$. Keeping just the leading
divergences and introducing $\rho = (\alpha\epsilon)^{5/4}$ the
metric expanded around $Q$ is \be \dd s^2 = \alpha\rho^{-2/5}(-\dd
T^2 + \dd \tilde{\Omega}_3^2) + \frac{16}{25} \dd \rho^2 +
\rho^{2/5}(\dd v_i^2 + \dd \Omega_3^2). \ee A short calculation
then shows that the metric is singular with scalar curvature as
$\rho \tend 0$ \be R = -\frac{5}{16\rho^2} \ee and the singularity
is clearly time-like with no horizon.

Singularities are located also on the subset of $\der\Sigma_0$
where $z\neq\pm\frac12$. All these singularities are naked and
null.

Indeed assuming that $1/4-z^2 \tend \alpha^2$ as $y \tend 0$ and
looking at the $\{t,y\}$ sections we find
\begin{equation}
  \dd s^2 = - \alpha^{-1}\, y\, \dd t^2 + \alpha\, y^{-1}\,\dd y.
\end{equation}
With the change of variables, $u=\sqrt{y/ \alpha}\,\e^{-t/2}$
$v=\sqrt{y/ \alpha}\,\e^{t/2}$, the metric becomes simply \be \dd
s^2 = \dd u \dd v \ee and the singularity is along the curves,
$u=0$ and $v=0$. The singularity is due to the way in which the
radii of the two three spheres, $S^3$ and $\tilde{S}^3$ go to zero
\cite{Myers:2001aq}.

\section{Singular solutions: some examples}\label{singular
solutions, general considerations on the examples}

Interpreting $\rho=1/2-z_0$ as the density of the dual
fermions, one first natural generalization of the boundary
conditions in \cite{bubbling} is to have density $\rho\neq1$. We
note that for generic $\rho(x_1,x_2)$, the radius of the
asymptotic $AdS_5\times S^5$ is given by
\begin{equation}\label{radius ads}
  R_{AdS}^4 = \frac 1\pi \int \rho(x_1,x_2)\dd^2 x
\end{equation}
We have that $0\leq\rho\leq 1$ (that is $-1/2\leq z_0\leq 1/2$),
if and only if $\der \hat \Sigma = \emptyset$. In this case all
the singularities will be null.

\begin{figure}
\begin{center}
\includegraphics[width=10 cm]{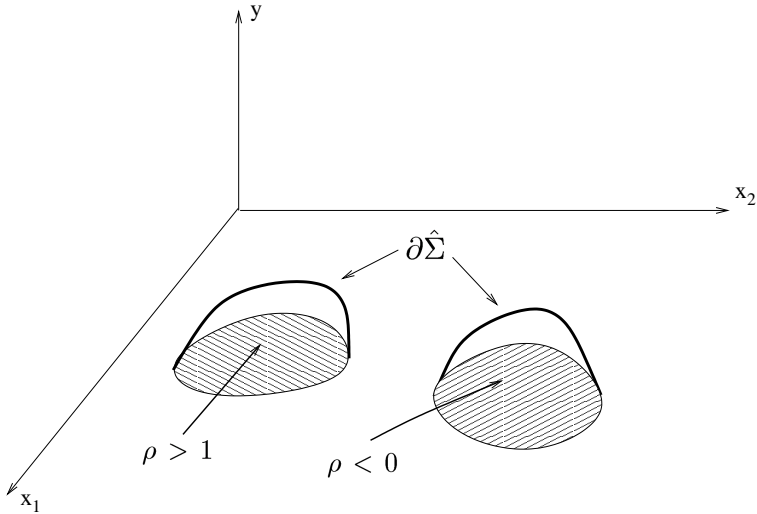}
\end{center}
\caption{Two regions of $y=0$ plane, one with $\rho>1$ and the
other with $\rho<0$, leading to a non trivial $\der\hat\Sigma$
attached to the $y=0$ plane}
\end{figure}

The mass of the excitation is now given by
\begin{equation}\label{generic mass of the excitation}
  M=\frac 1 {8\pi^2 l_P^8}\left[\frac 1
{2\pi}\int\rho(x_1,x_2) (x_1^2+x_2^2)\dd^2
    x-\left(\frac 1 {2\pi} \int \rho(x_1,x_2)\dd^2 x\right)^2\right]
\end{equation}
with origin chosen again in such a way that the dipole vanishes
\begin{equation}
\int_{\mathcal D}\rho(x_1,x_2)  x_i \dd^2 x = 0.
\end{equation}
We note that for fixed value of $R_{AdS}$ there is a lower bound
on the mass obtained for $\rho = \pi
R_{AdS}^4\delta^{(2)}(x_1,x_2)$,
\begin{equation}
  M_{\text{min}} =  - \frac {R_{AdS}^8} {32\pi^2 l_P^8}
\end{equation}

A priori we can consider also
$\rho(x_1,x_2)<0$ in some domains provided that the integral
defining $R_{AdS}^4$ remains positive.
One can easily see that the cases $\rho>1$ and $\rho<0$ correspond
to choosing a  $\der\hat\Sigma$ not empty and attached to the
$y=0$ plane, as in Figure 2.
Taking $\rho$ negative in some region we can easily obtain arbitrary large
negative value of the mass for fixed $R_{AdS}$. It's enough to
have $\rho<0$ even in a very small region provided it is located
at large $x_1^2+x_2^2$. In the next subsections we
will restrict to the case $\rho\geq0$, studying some examples
with features that will serve as a guide for the general analysis of
Section 4.

The appearance of CTCs, which we will show to be unavoidable in
Section 4, and unbounded from below negative mass values suggest
that one should consider as unphysical the geometries seeded by a
density $\rho$ that does not remain between 0 and 1. For the sake
of causality and for the stability of the quantum version of the
supergravity theory, these solutions should be regarded as
unphysical on the basis of some \emph{global} argument. If the
singularity was resolved by quantum effects through some local
mechanism and ``smoothed'', then the asymptotics and mass could
not change significantly; moreover, we know that the existence of
CTCs is a manifestation of global properties of the spacetime.
Before discussing the possibility of such a resolution we will
study these singular geometries in more detail.

For simplicity, we will first study the case of
piecewise constant $\rho$. Assuming
$\rho=\sum_i\beta_i\chi(\mathcal D_i)$ the $z$ function can be
written as
\begin{equation}
  z=1/2-\frac{y^2}\pi\sum_i\beta_i\int_{\mathcal
  D_i}\frac{\dd^2x'}{[(x_1-x_1')^2+(x_2-x_2')^2+y^2]^2}
\end{equation}
The solution seeded by these density distributions can have null singularities
or naked time machines. Solutions with null
singularities are already discussed in the literature in various
places
\cite{Myers:2001aq,Buchel:2004mc,Horava:2005pv,Gubser:2004xx,Bak:2005ef}
although we have some additional interesting observations to make.
These issues will be discussed in section 3.1. In section 3.2 and
3.3 we will discuss the general features of configurations with
naked timelike singularities, in particular illustrating a
novel geometric mechanism for producing CTCs.
In Section 3.4 we study the specific case of the geometry seeded
by a circular droplet of density $\beta>1$.

In Section \ref{lifting the sphere} we discuss a class of
solutions which does not have a density distribution
$\rho(x_1,x_2)$ in the $y=0$ plane as source, but rather appears
as a natural continuation of the solutions studied in \ref{the
circle}. These solutions are indeed determined by a
$\der\hat\Sigma$ which does not intersect the $y=0$ plane. They
exhibit CTCs and their mass is unbounded from below.  Any possible
direct connection to the free fermion picture is lost.

\subsection{$\beta_i\leq1$ with at least one $\beta_i < 1$}
This case was already briefly considered in \cite{bubbling}. These
geometries have null singularities located on the $y=0$ plane
inside the droplets. We will show in Section \ref{necessary
condition for CTCs} that also these geometries are free of CTCs.

It is straightforward to show that if $\beta_i\leq1$ the mass
given by \eqref{generic mass of the excitation} is always
nonnegative. These configurations can be viewed as an averaged
version of a dilute gas of fermions. In this case one can think
that the singularity is resolved by \emph{local} quantum effects
by the appearance of a ``spacetime foam"
\cite{Balasubramanian:2005kk} and in the dual theory by simply moving
to the Coulomb branch of the moduli space.

Geometries corresponding to a single circular droplet of density
$\beta<1$ are precisely the solutions considered in
\cite{Myers:2001aq}.

In the limit that the radius goes to infinity, this describes the
$N\tend \infty$ limit of the Coulomb branch in the dual gauge
theory, as amply discussed in \cite{bubbling}. The corresponding
classical geometry is singular but is regularized as above by the
dilute fermi gas, or in geometric language, a dilute gas of giant
gravitons, the geometry of which is clearly smooth.

This solution leads one to an interesting relation
between a limit of the dual SCFT and the singular
homogeneous plane wave metrics that arise generically
as the Penrose limit of ``reasonable'' space-time
singularities \cite{Blau:2004yi}.

For simplicity one can actually consider the boundary condition
$\rho = \beta<1$ for all $(x_1,x_2)$. Consider a null geodesic
that ends on the ``null'' singularity and take the Penrose Limit
with respect to this null geodesic.

In such a case it is easy to see that the resulting metric is
exactly, \be \dd s^2 = 2\dd u \dd v + (3 (x_1^2 + x_2^2) -
\sum_{i=1}^{6} w_i^2)\frac{\dd u^2}{u^2} + \dd x^2 + \dd w^2. \ee
In principle this provides a SYM dual description of the singular
plane waves as a limit (analogous to the BMN
\cite{Berenstein:2002jq} limit of AdS/CFT) of the $N\tend\infty$
Coulomb branch in the original dual CFT.

\subsection{Some $\beta_i>1$}
This boundary condition is equivalent to lifting  the surface
$z=-1/2$ above the $\{x_1,x_2\}$ plane keeping its boundary fixed
at $y=0$. The continuation of $z$ inside this surface to $y=0$
will give a non-trivial function everywhere less than $-1/2$. This
is the first example of the non-empty $\der \hat\Sigma$ introduced
in Section 2.

The emerging geometries have timelike singularities on
$\der\hat\Sigma$ and CTCs. They include also negative (but bounded
from below) mass excitations of $AdS_5\times S^5$, as anticipated
at the beginning of this section.

In the next subsections we will focus on the $\{x_1,x_2,y\}$
sections. They contain almost all of the interesting features.

\subsection{Zooming}\label{zooming}
We consider the leading term of the expansion of $z$ and $V$ for
points close to $y=0$ and the boundary of one droplet of constant
density $\beta>1$. More precisely, with $L$ the typical dimension
of the droplet and $R$ the radius of curvature of the boundary, we
assume that $y$ and the distance to the boundary are both much
smaller than  $L$ and $R$\footnote{For a calculation of the
subleading terms in such an expansion, see
\cite{Takayama:2005bc}.}. The leading term can be obtained solving
the equations for $z$ with boundary condition
\begin{equation}
  \rho(x_1,x_2)=\begin{cases}
    \beta \quad , &x_2<0\\
0 \quad, &x_2>0
  \end{cases}
\end{equation}
The case $\beta=1$ has already been considered in \cite{bubbling}
and corresponds to the maximally supersymmetric plane wave
\cite{Blau:2001ne}. We note here that only in the case of $\beta =1$ the
``zooming'' limit that we are considering here coincides with the Penrose
limit. Indeed the BFHP plane wave is \emph{the only} plane wave
geometry that can be obtained via the LLM construction and its
generalization with the most general boundary conditions on $z$
considered in Section \ref{LLM constr and generalization}. All
(generalized) LLM metrics, have 16 Killing spinors $\psi$ whose
bilinears $\bar\psi\Gamma^M\psi$ are null Killing
vectors\footnote{As all such bilinears in type IIB solutions
\cite{Hackett-Jones:2004yi}} \emph{but not} covariantly constant
(c.c.). Any plane wave has 16 Killing spinors with c.c. Killing
vector bilinear, and the only one which has 16 extra Killing
spinors is the maximally supersymmetric one. The details of the
proof can be found in the Appendix.

\begin{figure}
\begin{center}
\includegraphics[width=10 cm]{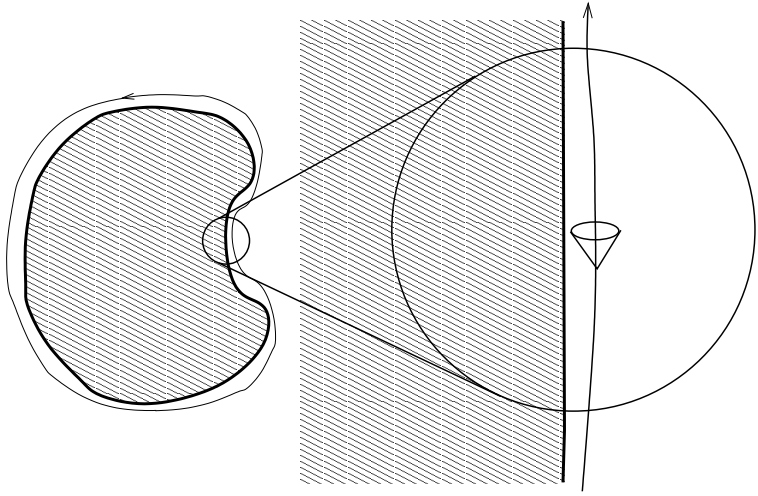}
\end{center}
\caption{Zoom showing light cone near a droplet with $\beta_i>1$
on the $\{x_1,x_2\}$ plane at $y=0$.}
\end{figure}

For generic $\beta$ we have
\begin{gather}
   z=\frac\beta2\frac{x_2}{\sqrt{x_2^2+y^2}}+\frac12(1-\beta)=
\frac\beta2 \cos\theta +
   \frac12(1-\beta)\\
   V_1=\frac{\beta}{2\sqrt{x_2^2+y^2}}=\frac\beta{2R}\qquad V_2=0
 \end{gather}
The plane $\cos\theta = \frac {\beta-2} \beta$ is $\der\hat\Sigma$
and the domain $\Sigma$ is defined by
\begin{equation}
1 >  \cos\theta > \frac{\beta-2}\beta
\end{equation}

The vector $\der_{x_1}$ is a Killing vector and
\begin{equation}
  g_{11}=\frac1{y\sqrt{\rho(1-\rho)}}\left(\frac\beta2(1-\beta)
(1-\cos\theta)\right)<0
\end{equation}
so that it is timelike. The limit $y\tend0,\cos\theta\tend1$ is
finite and gives
\begin{equation}
  g_{11}\tend(1-\beta)\sqrt\beta\frac1{2x_2}
\end{equation}
In the same limit, we have
\begin{equation}
  g_{22}=h^2\tend\frac{\sqrt\beta}{2x_2}
\end{equation}
In a neighborhood of $\der\mathcal D$, the $\{x_1,x_2\}$ plane is
thus a Lorentz submanifold. We note that the opening of the
lightcone is given by
\begin{equation}
  \tan \phi=\frac{\dd x_2}{\dd x_1}=\pm\sqrt{-\frac{g_{11}}{g_{22}}}=
\pm \sqrt{(\beta-1)}
\end{equation}

 From this analysis it is straightforward to conclude that if we
have a droplet $\mathcal D$ with $\beta_{\mathcal D}>1$ of smooth
boundary $\der\mathcal D$, \emph{ provided we stay close enough to
the $y=0$ plane and to $\der\mathcal D$ we have CTCs going around
$\mathcal D$} (Figure 3). Since these geometries have no horizon a
\emph{CTC passes through any point of the spacetime}.

\subsection{The disk}\label{the circle}

It is possible to perform a detailed analysis of the geometry
seeded by one single circular droplet of constant density
$\beta>1$. The analysis is interesting because it displays some generic
features of the timelike singular geometries and it is useful for
introducing the more general timelike singularities which we will
study in the next section.

We assume that the radius of the droplet is $R_0$. The radius of the
asymptotic $AdS_5\times S^5$ is thus given by
\begin{equation}
  R_{AdS}^4=\frac 1\pi \int\rho=\beta R_0^2
  \end{equation}
These geometries have already been studied in
\cite{Caldarelli:2004mz} where it is shown that they can be viewed
as a generalisation of the superstar studied in
\cite{Myers:2001aq}. The superstar geometries are parameterised by
a charge $Q$ and a scale parameter $L$, which are related to our
$\beta$ and $R_0$ in the following way.
\begin{equation}
  \beta=\frac1{1+Q/L^2}\,,\, R_0^2=L^2(L^2+Q)\,\then\, R_{AdS}=L
\end{equation}
For fixed value of $L$ we have
\begin{equation}
  -L^2<Q<0 \then \beta>1
\end{equation}
$Q=-L^2$ corresponds to $\rho=\pi L^4\delta^{(2)}(x_1,x_2)$. We
will discuss the continuation to $Q<-L^2$ in the next section.

Following the analysis in Section \ref{zooming} we expect to find
CTCs in these geometries. Going to polar
coordinates $R,\varphi$ in the $x_1,x_2$ plane we have
\begin{gather}
  z=\frac\beta2\frac{R^2-R_0^2+y^2}{\sqrt{(R^2+R_0^2+y^2)^2-4
  R^2 R_0^2}}+\frac 12 (1-\beta)\\
  V=V_\varphi\dd\varphi\\
  V_\varphi=\frac \beta2\left(1-\frac{R^2+R_0^2+y^2}{\sqrt{(R^2+R_0^2+y^2)^2-4
  R^2 R_0^2}}\right)
\end{gather}
The equation for the $\der \hat \Sigma$ is given by $z=-\frac12$
\begin{equation}
   R^2+\left(y-R_0
  \frac{\beta-2}{2\sqrt{\beta-1}}\right)^2 - R_0^2\frac {\beta^2}
{4(\beta-1)} =0\label{general value of rho}
\end{equation}
Thus the geometry is defined in the $y\geq0$ halfspace, outside a
sphere of radius $\frac\beta{2\sqrt{\beta-1}}R_0$ with centre at
$R=0$ and $y =\frac{\beta-2}{2\sqrt{\beta-1}}R_0$. In particular
it crosses the $y$ axis at $y=R_0\sqrt{\beta-1}$.

The square of the Killing vector $\der_\varphi$ is
\begin{equation}
  g_{\varphi\varphi}=-h^{-2}V_\varphi^2+R^2h^2
\end{equation}
and we have
\begin{equation}
  g_{\varphi\varphi}\geq0\iff
  \frac{y^2}{\beta-1}+\frac{R^2}{\beta}-R_0^2\geq0.
\end{equation}
The surface on which $g_{\varphi\varphi} =0$ is known as the
velocity of light surface (VLS).

\begin{figure}
\begin{center}
\includegraphics[width=10 cm]{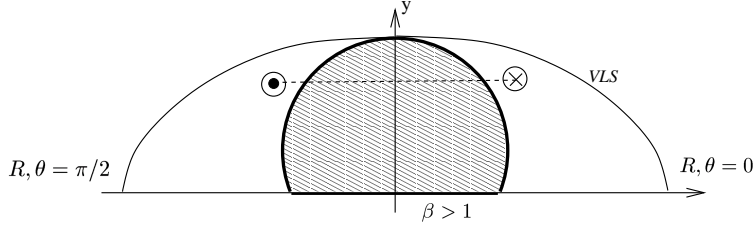}
\end{center}
\caption{Singularity and velocity of light surface for a disk with
$\beta>1$.}
\end{figure}

 From this analysis, three main features follow (see Figure 4). We will show in
Section \ref{sufficient condition} that they are generic for
geometries seeded by boundary conditions such that $\der \hat
\Sigma \neq\emptyset$.
\begin{enumerate}
\item  The VLS touches the
singularity where\footnote{Note that $V^2 = R^{-2} V_\varphi^2$
and $V_\varphi = O(R^2)$ as $R\tend0$} $V^2 = 0$. If the VLS did
not touch the singularity we would have CTCs which are
contractible to a point remaining timelike. At such a point the
local orientability of spacetime would be lost - possibly
indicating also a change in the signature of spacetime to two
time-like directions. The fact that the
VLS touches the singularity at a point, in such a way that there
is no loss of time orientability should be guaranteed, but we know
of no general theorem that proves this.

\item The opening of the lightcone in the $\{R,\varphi,y\}$
sections inside the ellipsoid is given by
\begin{equation}
  \tan\theta=\frac {y^2}{1/4-z^2} V_\varphi^2-R^2
\end{equation}
This means that provided we stay close enough to the singularity at
$z=-\frac12$ and that we go ``around'' it in the direction indicated
by $\der_\varphi$ we have CTCs.

\item All generalized LLM geometries are without horizon and thus
a CTC passes through any point of the spacetime.
\end{enumerate}

Following \eqref{generic mass of the excitation} we can calculate
the mass of these excitations over $AdS_5\times S^5$ as
\begin{multline}
  M=\frac1{8\pi^2 l_P^8}\left[\frac1\pi\int\rho\frac12
  (x_1^2+x_2^2)-\left(\frac1\pi\int\rho\frac12\right)^2\right]=\frac{(\beta
  R_0^2)^2}{32\pi^2 l_P^8}(\frac1\beta-1)
\end{multline}
Thus for $\beta>1$ a circular droplet seeds a negative mass
excitation. For fixed value of $R_{AdS}^4 = \beta R_0^2$, the
minimum mass is given by $M_{\text {min}}=-\frac{(\beta
  R_0^2)^2}{32\pi^2 l_P^8} = -\frac{R_{AdS}^8}{32\pi^2 l_P^8}$ and corresponds to $\beta=\infty$,
  $Q=-L^2$. As expected from the general considerations at the beginning of this Section,
   this corresponds to $\rho = \pi R_{AdS}^4
  \delta^{(2)}(x_1,x_2)$. In this case the surface $z=-1/2$ is a
  sphere of radius $R_{AdS}$, tangent to the
$\{x_1,x_2\}$ plane and centered on $(R,y)  =
  (0,\frac 12 R_{AdS})$. The VLS is determined by the saturation of
the inequality,
  \begin{equation}
    g_{\varphi\varphi}\geq 0 \iff y^2+R^2\geq R_{AdS}^2.
  \end{equation}

\subsection{Lifting the sphere}\label{lifting the sphere}
In the previous subsection we have considered geometries seeded by
a spherical $\der \hat \Sigma$ intersecting or tangent to the
$\{x_1,x_2\}$ plane. One could ask which geometries correspond to
a spherical $\der\hat\Sigma$ not touching the $\{x_1,x_2\}$ plane.
In this subsection we will answer this question. As in the case of
the circle of density $\beta>1$, these highly symmetric geometries
illustrate some features that will be shown to be generic for any
solution seeded by a $\der\hat\Sigma$ not attached to the
$\{x_1,x_2\}$ plane in Section \ref{lorentz topology}.

The functions
\begin{gather}
  z=\frac\beta2\frac{R^2-R_0^2+y^2}{\sqrt{(R^2+R_0^2+y^2)^2-4
  R^2 R_0^2}}+\frac 12 (1-\beta)\\
  V_\varphi=\frac \beta2\left(1-\frac{R^2+R_0^2+y^2}{\sqrt{(R^2+R_0^2+y^2)^2-4
  R^2 R_0^2}}\right)
\end{gather}
determine an asymptotically $AdS_5\times S^5$ provided $\beta
R_0^2>0$. Since $R_0^2$ (and not $R_0$) appears in these functions we
can analytically continue to $\beta<0$ and $R_0^2<0$. Recalling that
\begin{equation}
  \beta=\frac1{1+Q/L^2}\qquad R_0^2=L^2(L^2+Q)
\end{equation}
this corresponds to $Q<-L^2$.\\
We define for convenience
\begin{equation}
  \tilde R_0\equiv\sqrt{-R_0^2}
\end{equation}
and rewrite $z$ and $V$ as
\begin{gather}
   z=\frac\beta2\frac{R^2+\tilde R_0^2+y^2}{\sqrt{(R^2+y^2-\tilde R_0^2)^2+4
  R^2 \tilde R_0^2}}+\frac 12 (1-\beta)\label{z(R0tilde)}\\
V_\varphi=\frac \beta2\left(1-\frac{R^2+y^2-\tilde
R_0^2}{\sqrt{(R^2+y^2-\tilde R_0^2)^2+4
  R^2 \tilde R_0^2}}\right)\label{V(R0tilde)}
\end{gather}
This choice for $z$ corresponds to choosing $\der\hat\Sigma$ to be a
sphere of radius $\frac{-\beta}{2\sqrt{1-\beta}}\tilde R_0$ with
center at $R=0, y =\frac{2-\beta}{2\sqrt{1-\beta}}\tilde R_0$,
 $\der\Sigma_0$ coincides with the $\{x_1,x_2\}$ plane and
\begin{equation}
  z_0=\begin{cases}
    \frac12 &\text{on}\,\,\der\Sigma_0\\
    -\frac12 &\text{on}\,\,\der\hat\Sigma
  \end{cases}
\end{equation}
The expressions \eqref{z(R0tilde)},\eqref{V(R0tilde)} are the
analytic continuation of the solution for $z$ and $V$ with these
constraints. Clearly this continuation cannot be regular
everywhere inside the sphere and we expect to find a charge
somewhere. Looking at the leading order expansion of $\Phi$ for
$(R,y)=(R,\tilde R_0 + \eps)\tend(0,\tilde R_0)$
\begin{gather}
  \Phi=\frac {1/2-z} {y^2} \approx-\frac \beta {2\tilde R_0} \frac 1
  {\sqrt{R^2+\eps^2}}\\
  V_\varphi\approx \frac \beta 2 \left(1-\frac \eps
  {\sqrt{R^2+\eps^2}}\right)
\end{gather}
we can identify the charge and assume that $\Phi$ satisfies the
equation
\begin{equation}
  \ast_6\dd\ast_6\dd \Phi = 4\pi\frac\beta{2\tilde
  R_0}\,\delta(y-\tilde R_0)\,\delta ^{(2)}(R)
\end{equation}
We will briefly show in Section \ref{lorentz topology} that
whenever a subset of $\der \hat \Sigma$ is not attached to the
$\{x_1,x_2\}$ plane then we expect $\Phi$ to satisfy a similar
equation.

Integrating over the five-sphere at infinity we find that
\begin{equation}
  A=-\beta\tilde R_0^2=\beta R_0^2=L^4
\end{equation}
and so as expected $R_{AdS}=L$.\\
We have
\begin{equation}
g_{\varphi\varphi}\geq0\iff
  \frac{y^2}{1-\beta}+\frac{R^2}{-\beta}-\tilde R_0^2\geq0
\end{equation}
As happened in the case $-L^2<Q<0$ also here the velocity of light
surface touches the singularity, precisely at $R=0$ and $y=\tilde
R_0\sqrt{1-\beta}$. As already mentioned in that case we expect
this to be a general feature of geometries with CTCs and we will
show this in Section \ref{lorentz topology}. A more precise way to
state this situation is to say that inside the VLS the lightlike
direction has a non-trivial $\pi_1$.

On the segment of the $y$ axis, between the $y=0$ plane and the
lower intersection with the singularity at $y=\frac{\tilde
R_0}{\sqrt{1-\beta}}$ we have
\begin{equation}
  g_{\varphi\varphi}=- R_{AdS}^2 (-\beta)\frac1{\sqrt{\tilde
  R_0^2-(1-\beta)y^2}}
\end{equation}
Thus, the segment is actually a cylinder and so again there are no
CTCs which are contractible to a point while remaining timelike as
shown in Figure 5.

\begin{figure}
\begin{center}
\includegraphics[width=7 cm]{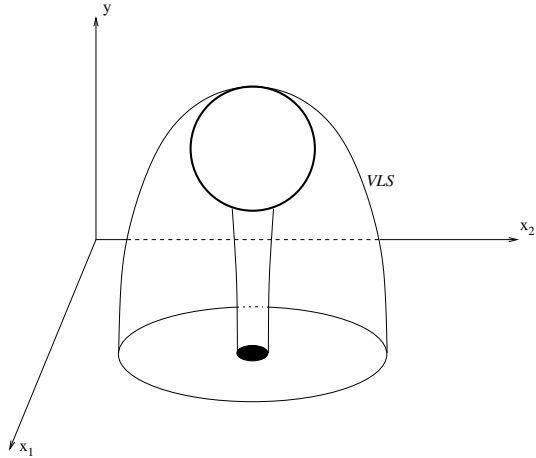}
\end{center}
\caption{``Medusa'' diagram: singularity, velocity of light
surface and cylinder connecting the singularity ($\der \hat \Sigma$)
to the $y=0$ plane, for the lifted sphere.}
\end{figure}

 Looking at the next to leading order
expansion of the metric for $R^2+y^2\tend\infty$ we can derive the
mass of these excitations of $AdS^5\times S^5$
\begin{equation}
  M=\frac{(\beta
  R_0^2)^2}{32\pi^2 l_P^8}(\frac1\beta-1)
\end{equation}
which is clearly negative and, for fixed $R_{AdS}$, tends to minus
infinity for $\beta\tend 0^-$.

\section{Singular solutions: generic properties}
In this section we will prove the following

\begin{teo}\label{necessary and sufficient condition for CTCs}
 Geometries of the type studied in Section 2 have
  closed timelike curves if and only if
$\der\hat\Sigma\neq\emptyset$
\end{teo}
In particular standard LLM geometries are free of CTCs as well as
all geometries seeded by boundary conditions such that
$\der\hat\Sigma=\emptyset$ and
\begin{equation}
  \frac 1 2 - z_0 (x_1,x_2) = \rho(x_1,x_2)\quad 0\leq\rho\leq1
\end{equation}
On the other hand, whenever $\rho>1$ or $\rho<0$, (and thus
$\der\hat\Sigma\neq\emptyset$), we have CTCs in the spacetime.

We will divide the proof into the 2 subsections 4.1 and 4.3. In
subsection 4.2 we will comment on the generic (Lorentz) topology
of the solutions and show that some of the interesting features of
the examples in Sections \ref{the circle} and \ref{lifting the
sphere} are indeed quite general.

\subsection{Sufficient condition for CTCs}\label{sufficient condition}
It's easy to show that when $\der\hat\Sigma\neq\emptyset$ we
have CTCs.\\
Looking at the asymptotic expansion for large values of
$x_1^2+x_2^2+y^2$ in Section \ref{Asymptotic behaviour}, we can
see that the vector field
\begin{equation}
  \der_\psi\equiv\frac 1 {\sqrt{V_1^2+V_2^2}} (V_1 \der_1 + V_2 \der_2)
\end{equation}
has closed\footnote{This is due to the gauge choice
$\der_1V_1+\der_2V_2=0$}, almost circular orbits at infinity. We
can shift $V$ by a constant amount such that $V=0$ at a point
$P\in\der\hat \Sigma$ with $\der_y z (P)\neq 0$ and the orbits of
$\der_\psi$ are closed around $P$. Let's assume for definiteness
that $z(P)=-\frac12$. In a neighborhood of P we have
\begin{gather}
  z(x_1,x_2,y)\approx-\frac12+\delta z\\
  V_i(x_1,x_2,y)\approx\delta V_i
\end{gather}
where $\delta z$ and $\delta V_i$ are linear in the
co-ordinates $(x_1-x(P),x_2-x(P),y-y(P))$. The metric of the sections
$\{x_1,x_2,y\}$ is (recalling that $h^4=\frac {1/4-z^2}{y^2}$),
\begin{equation}
\begin{split}
\tilde g &=
  \begin{pmatrix}
    h^2-h^{-2}V_1^2  & -h^{-2} V_1 V_2 & 0\\
    -h^{-2} V_1 V_2 & h^2-h^{-2}V_2^2 & 0\\
    0 & 0 & h^2
  \end{pmatrix}\approx\\&\approx \frac {1}{y(P)\sqrt{\delta z}}
  \begin{pmatrix}
    \delta z-y(P)^2\delta V_1^2 & y(P)^2\delta V_1 \delta V_2 &
    0\\
y(P)^2\delta V_1 \delta V_2 & \delta z-y(P)^2\delta V_2^2 &
0\\
0 & 0& \delta z
  \end{pmatrix}
\end{split}
\end{equation}
The vectors
\begin{equation}
  \begin{split}
    &\der_\psi \\
    &\der_\sigma\equiv\frac 1 {\sqrt{V_1^2+V_2^2}} (-V_2 \der_1 + V_1 \der_2)\\
    &\der_y
  \end{split}
\end{equation}
are eigenvectors of $\tilde g$ with eigenvalues respectively
\begin{equation}
  (h^2-h^{-2}(V_1^2+V_2^2)\,,\,h^2\,,\,h^2)\approx
  \frac{1}{y(P)\sqrt{\delta z}}(\delta z-y(P)^2\delta V^2\,,\, \delta
  z\,,\,
  \delta z)
\end{equation}
Thus for
\begin{equation}
  \delta z - y(P)^2 \delta V^2 < 0
\end{equation}
the sections are timelike. This equation also shows that the
velocity of light surface always touches the singularity where
$V=0$, as shown in Figure 6.

The opening of the lightcone is given by
\begin{equation}
  \tan\theta=h^{-4}(V_1^2+V_2^2)-1\approx \frac {y_(P)^2}{\delta
  z} \delta V^2-1
\end{equation}
Thus any closed curve going around $P$ in the sense indicated by
$\der_\psi$ is a CTC provided that we stay close enough to $P$ and
$\der\hat\Sigma$ (on which we recall $\delta z = 0$ by
definition). Since the CTCs are not hidden by a horizon, also in
this general case a CTC passes through any point of the spacetime.
\begin{figure}
\begin{center}
\includegraphics[width=10 cm]{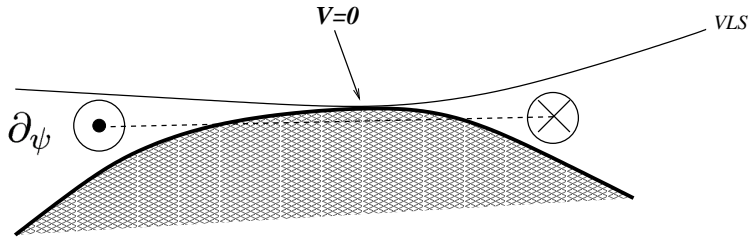}
\end{center}
\caption{$\der\hat \Sigma$ and the VLS touching at $V^2 = 0$.}
\end{figure}

\subsection{(Lorentz) topology}\label{lorentz topology}
In the case discussed in Section \ref{lifting the sphere} we have
$z_0=\frac12$ on the entire $\{x_1,x_2\}$ plane and $z=-\frac12$
on a sphere centered on the $y$ axis. The appearance of
contractible CTCs is excluded by a detailed analysis of the
structure of the metric. The Lorentz topology is thus nontrivial,
as one could expect in order to preserve the regularity of the
local structure of spacetime. The same analysis shows that the
topology of the $\{x_1,x_2,y\}$ sections is still $\RR^2\times
\RR^+$, even if at first sight one would say that a sphere has
been removed. This is essentially due to the non vanishing of
$V_\varphi$ along the $y$ axis in the segment between the $y=0$
plane and the sphere.

Assume we have a connected subset of $\der\hat\Sigma$ which is not
attached to the $\{x_1,x_2\}$ plane.  We can analytically continue
$z$ (and thus $\Phi$) to the $|z|>\frac 12 $ side of $\der\hat
\Sigma$. We will necessarily encounter some pole singularity in
the equation for $\Phi$, as $\delta$ sources centered on some
point $Q$. In a neighborhood of such a point $(\vec x_0,y_0)$ we
have to leading order
\begin{gather}
  z\approx \sigma y_0^2 \frac 1 {\sqrt{(y-y_0)^2+R^2}}\\
  V\approx V_\varphi\dd\varphi\qquad V_\varphi\approx\sigma
  y_0\left(1-\frac {y-y_0}{\sqrt{(y-y_0)^2+R^2}}\right)
\end{gather}
where $R,\varphi$ are polar coordinates in $x_1,x_2$ centered on
$\vec x_0$. By continuity, we can argue that in a neighborhood of
this $Q$, for $y<y_0$, the vector $V_i\der_i$ is circulating
around a line $\mathcal L$ on which it doesn't vanish. Going
locally to polar coordinates centered on the intersection of this
line with a constant $y$ plane, we have that
\begin{equation}
  g_{\varphi\varphi} = -h^{-2}V_\phi^2+R^2 h^2
\end{equation} is non vanishing at $R=0$ and thus the line $\mathcal L$ is topologically a cylinder. As in
section 3.5, the shape of the space-time around such a point $Q$
is similar to the ``Medusa'' diagram of Figure 5. We expect that
several disconnected components of $\der\hat\Sigma$ may give rise
to more complicated geometrical structures.

\subsection{Necessary condition for CTCs}\label{necessary condition for
CTCs} In this Section we will show that if $\der \hat \Sigma =
\emptyset$, then there are no CTCs. Looking at the metric
\eqref{geometry} it is clear that if the determinant $\tilde g$ of
the spatial section $\{x_1,x_2,y\}$ is positive, then there cannot
be CTCs. We recall from Section 2 that
\begin{gather}
  z(x_1,x_2,y)=\frac 1 2 - \frac {y^2} \pi \int \frac {\rho(x_1',x_2')\dd^2
  x}{\left[(x_1-x_1')^2+(x_2-x_2')^2+y^2\right]^2}\\
  V_i=-\frac1\pi\epsilon_{ij}\int\frac{(x_j-x_j')\rho(x_1',x_2')\dd^2
  x'}{[(x_1-x_1')^2+(x_2-x_2')^2+y^2]^2}\\
  \end{gather}
  and the determinant of the three dimensional sections
  \begin{multline}
  \tilde g = h^4-V^2=\frac {1/4-z^2}{y^2}-V^2=\\
  =\frac {1} \pi \int \frac {\rho(x_1',x_2')\dd^2
  x}{\left[(x_1-x_1')^2+(x_2-x_2')^2+y^2\right]^2}-\frac {y^2} {\pi^2}\left( \int \frac {\rho(x_1',x_2')\dd^2
  x}{\left[(x_1-x_1')^2+(x_2-x_2')^2+y^2\right]^2}\right)^2+\\-\sum_{i=1,2}\frac 1{\pi^2}\left(\int\frac{(x_i-x_i')\rho(x_1',x_2')\dd^2
  x'}{[(x_1-x_1')^2+(x_2-x_2')^2+y^2]^2}\right)^2\end{multline}
Any possible geometry seeded by a function $\rho(x_1,x_2)$ with
$0\leq\rho\leq1$ can be approximated as well as desired by a
piecewise constant $\bar\rho$ such that $\bar\rho=0,1$. So it is
enough to prove that the determinant is positive for standard LLM
geometries defined by droplets of density $\rho=1$.

We will prove that, given any possible distribution of droplets
$\mathcal D$ and any point $P\equiv(x_1(P),x_2(P),y)$, there is a
halfplane $\Pi$ distribution for which $z(P)$ is the same as for
the original distribution and $V(P)^2$ is larger. In this way the
determinant $\tilde g(P)_\Pi$ for the halfplane distribution is
smaller than the original determinant $\tilde g_{\mathcal D}$ . As
noted already in \cite{bubbling} a halfplane distribution
corresponds to the maximally supersymmetric plane wave and for this
metric the determinant always satisfies the relation
\begin{equation}
  \tilde g_\Pi= \frac{1/4-z^2}{y^2}- V_\Pi^2=0
\end{equation}
So we have $\tilde g_{\mathcal D}\geq\tilde g_\Pi=0$.

We first make some assumptions in order to simplify the proof.
Given the point $P\equiv(x_1(P),x_2(P),y)$ we move the origin of
the $\{x_1,x_2\}$ plane to $(x_1(P),x_2(P))$. We then define a
2-vector $\tilde V$ such that $\tilde V^2 = V^2 $
\begin{gather}
  \tilde V_i[\mathcal D] (P) =\frac1\pi\int_{\mathcal D}\frac{x_i}{\left(x_1^2+x_2^2+y^2\right)^2}\dd^2
  x\\
  \tilde V_1^2+\tilde V_2^2 =V_1^2+V_2^2
\end{gather}
where $\mathcal D$ is the union of all the droplets. We also have
\begin{equation}
  z_{\mathcal D}(P)=\frac 1 2 - \Delta_{\mathcal D}z=\frac 1 2 -\frac{y^2}\pi \int_{\mathcal D}\frac{1}{\left(x_1^2+x_2^2+y^2\right)^2}\dd^2
  x
\end{equation}

We identify the direction of $\tilde V$ with the $x_2$ axis. Let
us assume that the droplets are all contained in the strip
\begin{equation}
   x_{\text {min}}\leq x_2\leq x_{\text {max}}
\end{equation}
where one or even both of $x_{\text {min}}$ and $x_{\text {max}}$
can also be infinite. A distribution corresponding to the
(half)plane $\Pi_0$ defined by $x_2\geq x_{\text {min}}$ will give
us
\begin{equation}
  z_{\Pi_0}(P)\leq z_{\mathcal D}(P)
\end{equation}
since $\mathcal D \subseteq \Pi_0$

The equality holds just in the case that the original distribution is
already a halfplane\footnote{Or a completely filled plane, which
we neglect since it is trivial: the solution is empty Minkowski
space}. In all the other cases, we take a halfplane $\Pi$ defined
by
\begin{gather}
x_2\geq x
\end{gather}
with $x>x_{\text{min}}$ such that
\begin{equation}
  z_{\Pi}(P)=z_{\mathcal D}(P)
\end{equation}

\begin{figure}
\begin{center}
\includegraphics[width=7 cm]{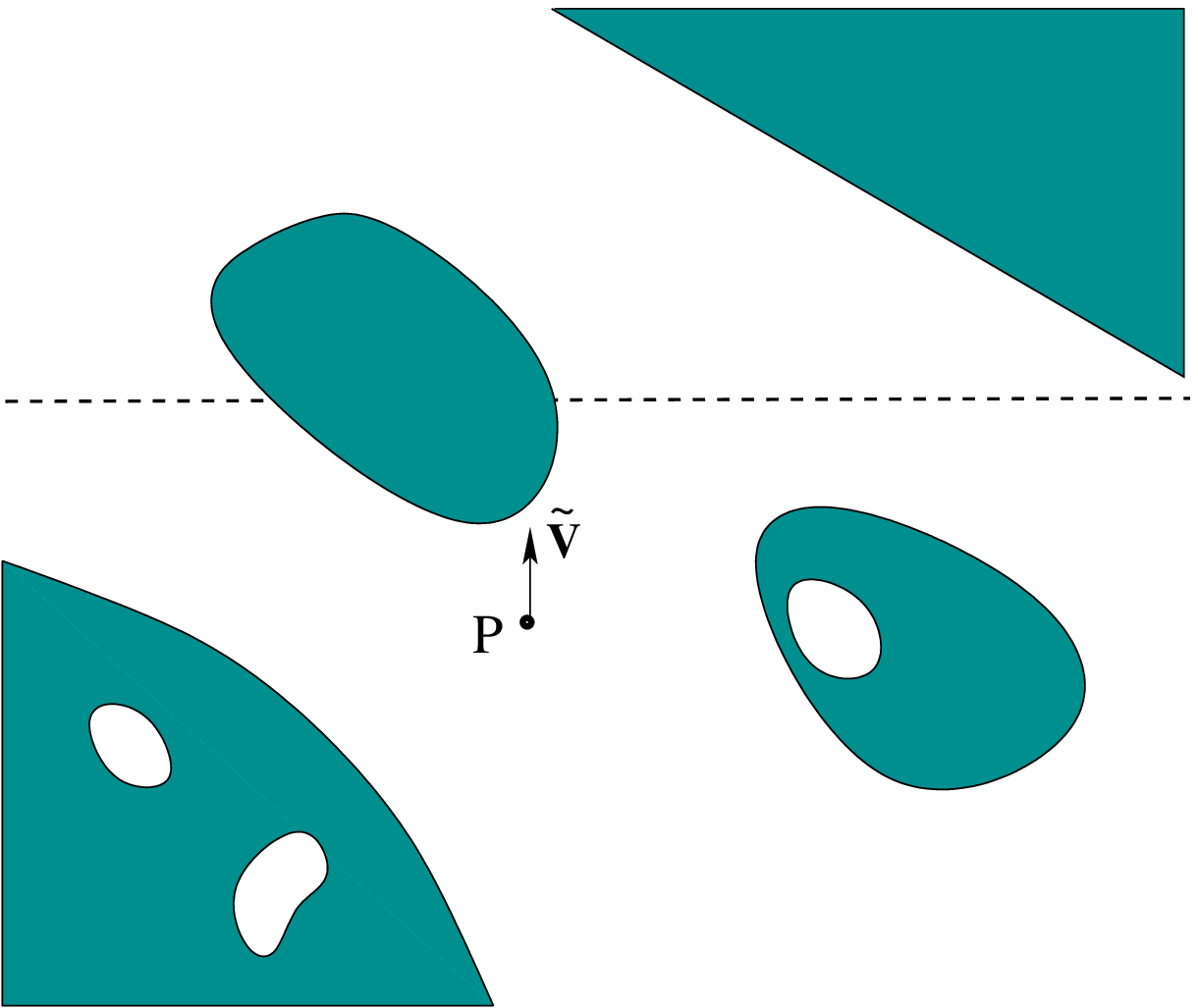}
\hfill
\includegraphics[width=7 cm]{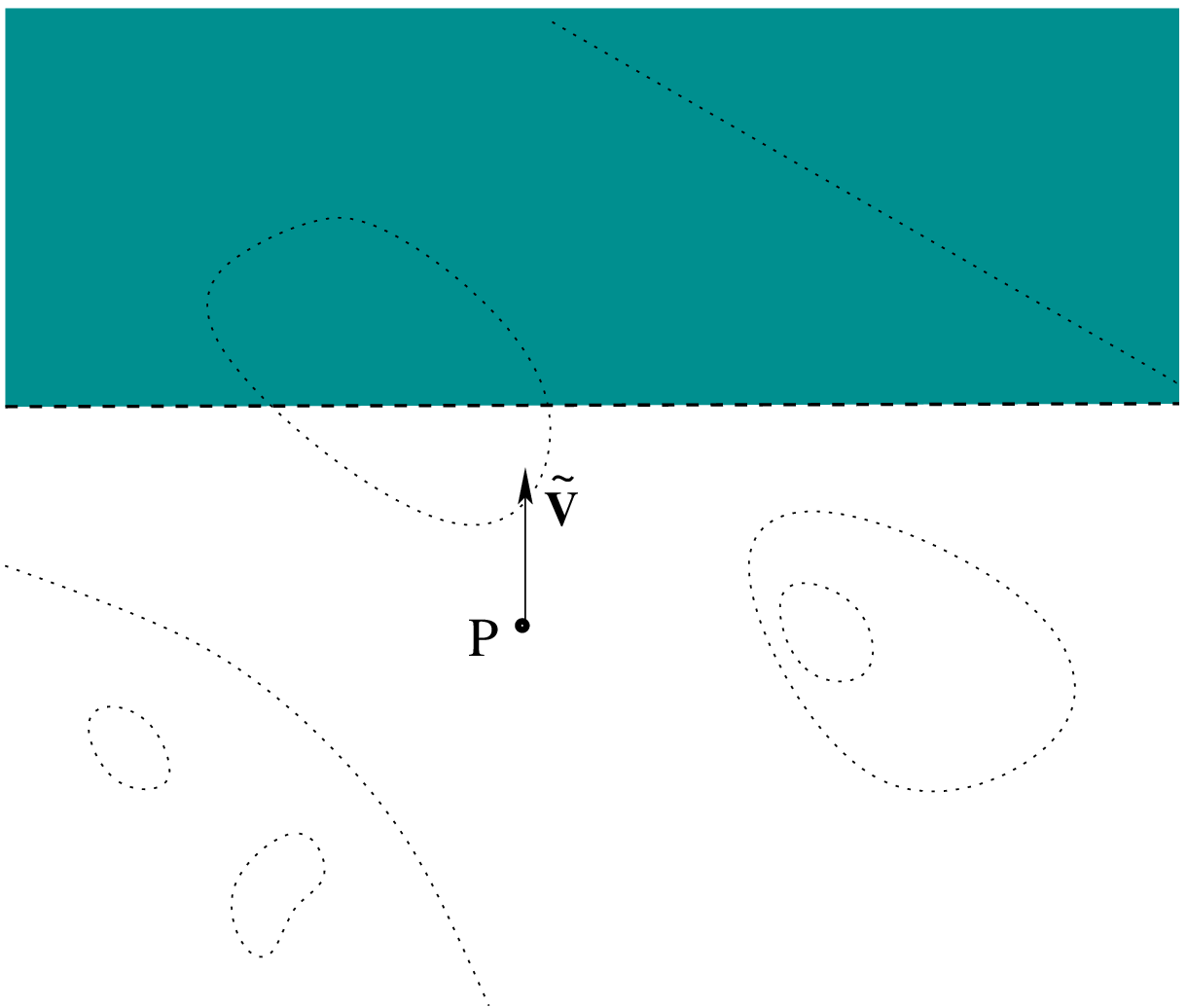}
\end{center}
\caption{Changing $\mathcal D$ into $\Pi$}
\end{figure}

We note that for a generic domain $\mathcal D$ we have
 the following relation between $\Delta_{\mathcal D} z$
and  $\tilde V_2[\mathcal D](P)$
\begin{equation}\label{Delta V2 and Delta z}
  \tilde V_2[\mathcal D](P) = \frac{\Delta_{\mathcal D} z}{y^2} \langle x_2 \rangle_{\mathcal D}
 \end{equation}
with
\begin{gather}
  \langle x_2\rangle_{\mathcal D} = \frac 1 \pi \int_{\mathcal D}
  \frac{x_2\dd^2x}{\left(x_1^2+x_2^2+y^2\right)^2}\left(\frac 1 \pi \int_{\mathcal D}
  \frac{\dd^2x}{\left(x_1^2+x_2^2+y^2\right)^2}\right)^{-1}=\\\notag
  =\int x_2\,\,\mu_{\mathcal D}(x_2)\dd x_2\\
  \mu_{\mathcal D}(x_2)=\frac 1 \pi \int_{\mathcal D}
  \frac{\dd x_1 }{\left(x_1^2+x_2^2+y^2\right)^2}\left(\frac 1 \pi \int_{\mathcal D}
  \frac{\dd x_1\dd x_2}{\left(x_1^2+x_2^2+y^2\right)^2}\right)^{-1}\\
  \int \,\,\mu_{\mathcal D}(x_2)\dd
  x_2= 1
\end{gather}
Thus $\mu_{\mathcal D}(x_2)$ acts as a normalized weight function.

 From the definition of $\mu_{\mathcal D} (x_2)$ and from the fact
that, by definition of $\Pi$
\begin{gather}\label{equality of Delta z}
  \Delta_{\mathcal D}z = \Delta_{\Pi }z\\\notag
  \text{i.e.}\\
  \frac{1}\pi \int_{\mathcal D}\frac{1}{\left(x_1^2+x_2^2+y^2\right)^2}\dd^2
  x = \frac{1}\pi \int_{\Pi}\frac{1}{\left(x_1^2+x_2^2+y^2\right)^2}\dd^2
  x
\end{gather}
one can easily see that
\begin{gather}
  \mu_\Pi (x_2) \geq \mu_{\mathcal D} (x_2)  \quad,\quad x_2\geq
  x\\
  \mu_\Pi(x_2) = 0 \quad , \quad x_2<x
\end{gather}
We have
\begin{multline}
  \langle x_2\rangle_\Pi = x + \langle (x_2-x)\rangle_\Pi = \\
  =x+\int\mu_\Pi(x_2)(x_2-x)\dd x_2\geq x + \int_{x_2>x}\mu_{\mathcal D}(x_2)(x_2-x)\dd
  x_2\,>\,\langle x_2\rangle_{\mathcal D}
\end{multline}
The last inequality holds because
\begin{multline}
  \langle x_2\rangle_{\mathcal D} = x + \int\mu_{\mathcal D}(x_2)(x_2-x)\dd
  x_2 = \\
  =x + \int_{x_2>x}\mu_{\mathcal D}(x_2)(x_2-x)\dd
  x_2 +  \int_{x_2<x}\mu_{\mathcal D}(x_2)(x_2-x)\dd
  x_2
\end{multline}
and the last term is clearly negative.

Recalling \eqref{Delta V2 and Delta z} and \eqref{equality of
Delta z} we conclude
\begin{equation}
  \tilde V_2[\Pi](P)>\tilde V_2 [\mathcal D](P)
\end{equation}
and thus we have
\begin{equation}
  \tilde g_{\mathcal D}=\frac{1/4-z_{\mathcal D}(P)^2}{y^2}-V_{\mathcal D}^2
  >
  \frac{1/4-z_{\Pi}(P)^2}{y^2}-V_{\Pi}^2=0
\end{equation}

\subsubsection*{The case $y=0$}
In the proof we have implicitly assumed $y>0$. In the limit
$y\tend 0$ one can argue, by continuity
\begin{equation}
  \tilde g_{\mathcal D} \geq 0
\end{equation}
With a bit of effort, we can prove that the equality holds only
for the halfplane.

Instead of choosing $x$ in order to fix $z(P)$ we decide to fix
\begin{equation}
\lim_{y\tend 0}  \frac {1/4-z^2}{y^2}(P)
\end{equation}
which is finite since by
hypotheses $z\tend\pm\frac12$ and is even in $y$. Recalling that
\begin{equation}
   \lim_{y\tend 0} \frac {y^2} \pi \int_{\mathcal D}
  \frac{\dd^2x}{\left(x_1^2+x_2^2+y^2\right)^2} = \begin{cases}
    1 &P\in\mathcal D\\
    0 &P\in \bar{\mathcal D}
  \end{cases}
  \end{equation}
we have
\begin{equation}
    \frac {1/4-z^2}{y^2}\tend \begin{cases}
      \frac {1}{y^2} \Delta_{\mathcal D} z
  \tend \frac 1 \pi \int_{\mathcal D}
  \frac{\dd^2x}{\left(x_1^2+x_2^2\right)^2} & P\in \bar {\mathcal
  D}\\
 -\frac {1}{y^2}\Delta_{\bar{\mathcal D}} z
  \tend -\frac 1 \pi \int_{\bar{\mathcal D}}
  \frac{\dd^2x}{\left(x_1^2+x_2^2\right)^2} & P\in \mathcal
  D
    \end{cases}
  \end{equation}
Noting that $V_{\mathcal D} = - V_{\bar{\mathcal D}}$, in both
cases we can use the same argument as for $y\neq 0$ provided that
we change $\mathcal D$ into $\bar{\mathcal D}$ when $P\in
{\mathcal D}$. Thus $\tilde g\geq0$ and again the equality holds
only for the halfplane.

\section{Supergravity singularities and dual field theories}

There already exist in the literature on $AdS/CFT$ duality, some
indications that geometries with naked time machines are related
to non-physical phenomenon in the dual gauge theory. The dual
picture should provide a field theory interpretation for the
quantum mechanism at work in the resolution of these pathologies,
possibly through a careful treatment of unitarity.

In particular, the overrotating solutions of
\cite{Breckenridge:1996is} are exactly of this type and as already
noted in that paper, and further elucidated in
\cite{Gibbons:1999uv,Herdeiro:2000ap}, the operator in the
corresponding D-brane configuration that takes an underrotating
geometry to an overrotating one is non-unitary.

In that case it was first noticed \cite{Gibbons:1999uv} that the
overrotating geometries have a VLS that repulses all geodesics
that approach from the outside, and thus the region of CTCs is
effectively removed from the space-time. It was then noticed in a
series of works on the enhancon mechanism that incorporating extra
charge sources one can remove the causality violating
region\cite{Jarv:2002wu}. A similar idea is developed for example
also in \cite{Drukker:2004zm}. Our naked time machines do not have
a repulsive VLS and as a consequence this method for removing the
singularity cannot be applied here.

That some form of chronology protection mechanism should however
be present has been conjectured in \cite{Caldarelli:2004mz}. In
this paper the rotationally symmetric singular configurations that
we have studied in Section \ref{the circle} are noted to not have
a description in terms of the dual free fermion picture as they
violate the Pauli exclusion principle.

In general relativity and in supergravity there are of course many
geometries that contain CTCs and naked singularities. Is it
possible that a similar principle could also rule out those
geometries? In particular is it possible that these geometries are
in general related to non-unitarity in the dual gauge theories?
The violation of the Pauli exclusion principle suggests that our
naked time machines may more generally be related to some
non-unitary behaviour in the dual gauge theory\footnote{For a
recent and somewhat different perspective on the relationship
between unitarity and CTCs, see \cite{Cornalba:2005je}.}.

The conformal dimension $\Delta$ of an operator in the $CFT$ dual
to an asymptotically $AdS_5\times S^5$ geometry is equal to the
mass or angular momentum ($M=J$ as a consequence of the BPS
condition) of the configuration. For a solution seeded by a
density distribution $\rho$
\begin{equation}
 \Delta=M=\frac1{8\pi^2 l_P^8}\left[\frac1\pi\int\rho\frac12
  (x_1^2+x_2^2)-\left(\frac1\pi\int\rho\frac12\right)^2\right]
\end{equation}

As noted in Section 3.4, for a density which is $\beta$ inside a
disk, we have
\begin{equation}\label{mass and conformal dimension of a disk}
   M = \Delta = \frac{\beta
   R_{AdS}^8}{32\pi^2 l_P^8}(\frac1\beta-1)
\end{equation}
From the CFT point of view a configuration with $\beta \gtrsim 1$
can be seen as a ``small'' deformation of a configuration with
$\beta = 1$ and slightly larger radius. Equation \eqref{mass and
conformal dimension of a disk} shows that this deformation
corresponds to an operator with {\it negative} conformal
dimension.

In general we expect, even though we cannot prove it directly,
that configurations with $\rho$ not between 0 and 1 correspond to
deformations of the $CFT$ by negative conformal dimension
operators. As seen in Section 3.5, solutions with more general
boundary conditions can still be interpreted as continuous
deformations of solutions seeded by density distributions and a
similar argument should also relate them to operators of negative
conformal dimension.

In a series of papers \cite{dobrev} all unitary irreducible
representations of the relevant superconformal algebra,
$su(2,2|N)$, are found and in particular unitarity requires that
they have positive conformal dimension. The ``unphysical''
geometries that we have studied in this paper then apparently
correspond to deformations by non-unitary operators (with negative
conformal dimension) in the dual $CFT$. This observation together
with the observed violation of the Pauli exclusion principle
provides strong evidence for the existence of a theorem, for $1/2$
BPS configurations in IIB supergravity, relating the chronology
protection conjecture to unitarity in the dual $CFT$.

\section*{Acknowledgments}
We would like to thank V. Balasubramanian, M. Blau, M. Caldarelli,
E. Gava, K.S. Narain, M.S. Narasimhan, G. Thompson and N.
Visciglia for useful discussions and suggestions, and to the
organizers of the String Cosmology Workshop held in Uppsala,
Sweden, for the opportunity to present this work while it was
still in progress.

This research is supported by the Italian MIUR under the program,
``Teoria dei Campi, Superstringhe e Gravit\`a''.

\appendix

\section{Plane wave solutions}
Given any supersymmetric solution in supergravity, and in
particular in type IIB, one can construct a Killing vector
$\kappa$ by forming bilinears of the Killing spinors $\psi$
\begin{equation}
  \kappa^M =  \bar\psi\Gamma^M\psi
\end{equation}
In type IIB supergravity, we have $\kappa^M \kappa_M = 0$
\cite{Hackett-Jones:2004yi}.

For the geometry studied in \cite{bubbling} and in this paper the
timelike Killing vector $\der_t$ is obtained by adding one of
these bilinears to a Killing vector coming from the $SO(4)\times
SO(4)$ symmetry. We will briefly show that such timelike Killing
vector can be built only when the bilinear $\kappa$ is not
covariantly constant (c.c.). Any plane wave geometry in type IIB
has 16 Killing spinors whose bilinears are constant multiple of
the c.c. Killing vector of the metric \cite{Blau:2002mw}. If a
plane wave is to be in the class of solutions constructed in
\cite{bubbling}, it will have 16 extra Killing spinors and thus
must be the maximally supersymmetric pane wave studied in
\cite{Blau:2001ne}.

We refer to Appendix A of \cite{bubbling} for notation and
conventions.

\subsection{Ansatz and basic assumptions}
The supersymmetric type IIB solutions under examination are
described by
\begin{gather}
  \dd s^2=g_{\mu\nu}\dd x^\mu
  \dd x^\nu+\e^{H+G}\dd\Omega^2_3+\e^{H-G}\dd\widetilde{\Omega}_3^2\label{genericansatz}\\
  F_{(5)}=F_{\mu\nu}\dd x^\mu\wedge\dd x^\nu\wedge\dd\Omega+\widetilde{F}_{\mu\nu}\dd x^\mu\wedge\dd
  x^\nu\wedge\dd\widetilde{\Omega}
\end{gather}
A supersymmetric supergravity solution with just the $F_5$ field
strength turned on is characterized by a non vanishing 32
dimensional complex spinor $\eta$ satisfying
\begin{equation}\label{killspineq}
  \nabla_M\eta+\frac\ii
  {480}\Gamma_{M_1M_2M_3M_4M_5}F_{(5)}^{M_1M_2M_3M_4M_5}\Gamma_M\eta
  = 0
\end{equation}
Due to our symmetry assumptions, the generic solution can be
written as
\begin{equation}
  \eta=\eps_{a,b}\otimes\chi_a\otimes\tilde\chi_b
\end{equation}
with $\eps_{a,b}$ an 8 dimensional spinor and $\chi_a$,$\chi_b$ 2
dimensional spinors obeying the Killing spinor equation on the
Euclidean 3-sphere
\begin{equation}\label{spherekilling}
  \bar\nabla_c\chi=\alpha\frac\ii2\sigma_c \chi
\end{equation}
 with
$\alpha=a,b$ respectively, $\sigma_c$ in the Clifford algebra of
$SO(3)$ and $\bar\nabla_c$ the standard covariant derivative on
the Euclidean 3-sphere. Integrability conditions imply
$\alpha=\pm1$. There are $2$ linearly independent solutions for
each value of $\alpha$ \cite{Blau:2000xg,barkilling}.

\subsection{Spinor bilinears}
We define the set of spinor bilinears
\begin{equation}
  \begin{split}
    &K_\mu=-\bar\eps\gamma_\mu\eps\qquad f_1=\ii\bar\eps\sigma_1\eps\qquad f_2=\ii\bar\eps\sigma_2\eps\\
    &\bar\eps=\eps^\dag\gamma_0
  \end{split}
\end{equation}
One can show that
\begin{gather}\label{nabla mu K nu}
  \nabla_\mu
  K_\nu=\e^{-\frac32(H+G)}F_{\mu\nu}f_2+\e^{-\frac32(H-G)}\tilde
  F_{\mu\nu}f_1
\end{gather}
from which we can see that $K_\mu$ is a Killing vector for
$g_{\mu\nu}$ and by Fierz rearrengment one can show that
\begin{equation}\label{K^2norm}
  K^2=-f_1^2-f_2^2
\end{equation}
The standard ten dimensional Killing vector coming from the
sandwich of the spinors is given by, in ten dimensional covariant
tangent space components
\begin{equation*}
  \kappa=\left(K(\chi^\dag\chi)(\tilde\chi^\dag\tilde\chi),f_2\chi^\dag\vec{\sigma}\chi(\tilde\chi^\dag\tilde\chi),-f_1(\chi^\dag\chi)\tilde\chi^\dag\vec\sigma
\tilde\chi\right)
\end{equation*}
As expected from general considerations
\begin{multline}
  \kappa^2=K^2(\chi^\dag\chi)^2(\tilde\chi^\dag\tilde\chi)^2+f_2^2\sum_a(\chi^\dag\sigma_a\chi)(\chi^\dag\sigma_a\chi)(\tilde\chi^\dag\tilde\chi)^2+
  f_1^2(\chi^\dag\chi)^2\sum_a(\tilde\chi^\dag\sigma_a\tilde\chi)(\tilde\chi^\dag\sigma_a\tilde\chi)=\\
  =\left(-f_1^2-f_2^2+f_2^2+f_1^2\right)(\chi^\dag\chi)^2(\tilde\chi^\dag\tilde\chi)^2=0
\end{multline}
where we have used \eqref{K^2norm} and the basic fact, true for
every two dimensional spinor $\zeta$ that
\begin{equation*}
  \sum_a(\zeta^\dag\sigma_a\zeta)(\zeta^\dag\sigma_a\zeta)=(\zeta^\dag\zeta)^2
\end{equation*}
 We also have that
\begin{equation}
  \der_{\hat a}(\chi^\dag\chi)=\der_{\tilde
  a}(\tilde\chi^\dag\tilde\chi)=0
\end{equation}
and the vectors
\begin{equation}
  J\equiv(\vb 0,\e^{\frac12(H+G)}\chi^\dag\vec\sigma\chi,\vec0)\quad,\quad \tilde J\equiv(\vb 0,\vec0,\e^{\frac12(H-G)}\tilde\chi^\dag\vec\sigma\tilde\chi)
\end{equation}
are Killing vectors, corresponding to the $SO(4)\times SO(4)$
isometry of our ansatz.

\subsection{Analysis of bilinears}
 Assume that we have $f_{1,2}\neq0$. In this case the vector $\kappa$ is Killing but not covariantly constant (c.c.).
 In \cite{bubbling} it is shown that we have
\begin{equation}
  f_{1,2}\propto \e^{\frac12(H\mp G)}
\end{equation}
and thus also the vector
\begin{equation}
  \kappa+J(\tilde\chi^\dag\tilde\chi)+\tilde J (\chi^\dag\chi) = (K,\vec0,\vec0)
\end{equation}
obtained as the sum of $\kappa$ and a Killing vector of the
$SO(4)\times SO(4)$ symmetry, is a Killing vector for the full
metric.

This vector is identified with $\der_t$, which is possible since
\begin{equation}
  K^2 = -f_1^2-f_2^2<0
\end{equation}

 The fact that $f_{1,2}\neq0$ is thus crucial for all the
construction of the 1/2 supersymmetric solutions in
\cite{bubbling}. From \eqref{nabla mu K nu} one can see that the
16 independent Killing spinors $\eta$ do not have c.c. vector
bilinear. Any plane wave geometry in type IIB has 16 Killing
spinors whose bilinears are constant multiple of the c.c. Killing
vector of the metric. The vector bilinears coming from these
Killing spinors would then be null and c.c., with
\begin{equation}
 f_1=f_2=0
\end{equation}
which leads to
\begin{equation*}
  \kappa=(K(\chi^\dag\chi)(\tilde\chi^\dag\tilde\chi),\vec0,\vec0)
\end{equation*}
and $K^2 = 0$. In this case, by a similar construction to that of
LLM  \cite{ourconstructionforplanewave}, one can obtain a set of
plane waves with $SO(4)\times SO(4)$ isometry and non vanishing
five-form.

If one of these plane wave solutions could be obtained with the
techniques presented in \cite{bubbling} it should have 16 extra
Killing spinors whose bilinears $f_{1,2}$ do not vanish. This
means that the solution must have 32 Killing spinors and then
clearly it is the maximally supersymmetric plane wave of
\cite{Blau:2001ne}.

\providecommand{\href}[2]{#2}\begingroup\raggedright\endgroup


\begin{thebibliography}{10}

\bibitem{bubbling}
H.~Lin, O.~Lunin, and J.~Maldacena, ``Bubbling AdS space and 1/2
BPS
  geometries,''
\href{http://www.arXiv.org/abs/hep-th/0409174}{{\tt
hep-th/0409174}}.
%%CITATION = HEP-TH 0409174;%%.

\bibitem{Liu:2004hy}
J.~T. Liu and D.~Vaman, ``Bubbling 1/2 BPS solutions of minimal
six-dimensional
  supergravity,''
\href{http://www.arXiv.org/abs/hep-th/0412242}{{\tt
hep-th/0412242}}.
%%CITATION = HEP-TH 0412242;%%.

\bibitem{Liu:2004ru}
J.~T. Liu, D.~Vaman, and W.~Y. Wen, ``Bubbling 1/4 BPS solutions
in type IIB
  and supergravity reductions on S**n x S**n,''
\href{http://www.arXiv.org/abs/hep-th/0412043}{{\tt
hep-th/0412043}}.
%%CITATION = HEP-TH 0412043;%%.

\bibitem{Martelli:2004xq}
D.~Martelli and J.~F. Morales, ``Bubbling AdS(3),'' {\em JHEP}
{\bf 02} (2005)
  048,
\href{http://www.arXiv.org/abs/hep-th/0412136}{{\tt
hep-th/0412136}}.
%%CITATION = HEP-TH 0412136;%%.

\bibitem{Chong:2004ce}
Z.~W. Chong, H.~Lu, and C.~N. Pope, ``BPS geometries and AdS
bubbles,'' {\em
  Phys. Lett.} {\bf B614} (2005) 96--103,
\href{http://www.arXiv.org/abs/hep-th/0412221}{{\tt
hep-th/0412221}}.
%%CITATION = HEP-TH 0412221;%%.

\bibitem{Ghodsi:2005ks}
A.~Ghodsi, A.~E. Mosaffa, O.~Saremi, and M.~M. Sheikh-Jabbari,
``LLL vs. LLM:
  Half BPS sector of N = 4 SYM equals to quantum Hall system,''
\href{http://www.arXiv.org/abs/hep-th/0505129}{{\tt
hep-th/0505129}}.
%%CITATION = HEP-TH 0505129;%%.

\bibitem{Ebrahim:2005uz}
H.~Ebrahim and A.~E. Mosaffa, ``Semiclassical string solutions on
1/2 BPS
  geometries,'' {\em JHEP} {\bf 01} (2005) 050,
\href{http://www.arXiv.org/abs/hep-th/0501072}{{\tt
hep-th/0501072}}.
%%CITATION = HEP-TH 0501072;%%.

\bibitem{Sheikh-Jabbari:2005mf}
M.~M. Sheikh-Jabbari and M.~Torabian, ``Classification of all 1/2
BPS solutions
  of the tiny graviton matrix theory,'' {\em JHEP} {\bf 04} (2005) 001,
\href{http://www.arXiv.org/abs/hep-th/0501001}{{\tt
hep-th/0501001}}.
%%CITATION = HEP-TH 0501001;%%.

\bibitem{Mukhi:2005cv}
S.~Mukhi and M.~Smedback, ``Bubbling orientifolds,''
\href{http://www.arXiv.org/abs/hep-th/0506059}{{\tt
hep-th/0506059}}.
%%CITATION = HEP-TH 0506059;%%.

\bibitem{Spalinski:2005ha}
M.~Spalinski, ``Some half-BPS solutions of M-theory,''
\href{http://www.arXiv.org/abs/hep-th/0506247}{{\tt
hep-th/0506247}}.
%%CITATION = HEP-TH 0506247;%%.

\bibitem{Bena:2005va}
I.~Bena and N.~P. Warner, ``Bubbling supertubes and foaming black
holes,'' \href{http://www.arXiv.org/abs/hep-th/0505166}{{\tt
hep-th/0505166}}.
%%CITATION = HEP-TH 0505166;%%.

\bibitem{Mandal:2005wv}
G.~Mandal, ``Fermions from half-BPS supergravity,''
\href{http://www.arXiv.org/abs/hep-th/0502104}{{\tt
hep-th/0502104}}.
%%CITATION = HEP-TH 0502104;%%.

\bibitem{Grant:2005qc}
L.~Grant, L.~Maoz, J.~Marsano, K.~Papadodimas, and V.~S. Rychkov,
  ``Minisuperspace quantization of 'bubbling AdS' and free fermion droplets,''
\href{http://www.arXiv.org/abs/hep-th/0505079}{{\tt
hep-th/0505079}}.
%%CITATION = HEP-TH 0505079;%%.

\bibitem{Balasubramanian:2004nb}
V.~Balasubramanian, D.~Berenstein, B.~Feng, and M.-x. Huang,
``D-branes in
  Yang-Mills theory and emergent gauge symmetry,'' {\em JHEP} {\bf 03} (2005)
  006,
\href{http://www.arXiv.org/abs/hep-th/0411205}{{\tt
hep-th/0411205}}.
%%CITATION = HEP-TH 0411205;%%.

\bibitem{Berenstein:2004kk}
D.~Berenstein, ``A toy model for the AdS/CFT correspondence,''
{\em JHEP} {\bf
  07} (2004) 018,
\href{http://www.arXiv.org/abs/hep-th/0403110}{{\tt
hep-th/0403110}}.
%%CITATION = HEP-TH 0403110;%%.

\bibitem{Corley:2001zk}
S.~Corley, A.~Jevicki, and S.~Ramgoolam, ``Exact correlators of
giant gravitons
  from dual N = 4 SYM theory,'' {\em Adv. Theor. Math. Phys.} {\bf 5} (2002)
  809--839,
\href{http://www.arXiv.org/abs/hep-th/0111222}{{\tt
hep-th/0111222}}.
%%CITATION = HEP-TH 0111222;%%.

\bibitem{Myers:2001aq}
R.~C. Myers and O.~Tafjord, ``Superstars and giant gravitons,''
{\em JHEP} {\bf
  11} (2001) 009,
\href{http://www.arXiv.org/abs/hep-th/0109127}{{\tt
hep-th/0109127}}.
%%CITATION = HEP-TH 0109127;%%.

\bibitem{Buchel:2004mc}
A.~Buchel, ``Coarse-graining 1/2 BPS geometries of type IIB
supergravity,''
\href{http://www.arXiv.org/abs/hep-th/0409271}{{\tt
hep-th/0409271}}.
%%CITATION = HEP-TH 0409271;%%.

\bibitem{Horava:2005pv}
P.~Horava and P.~G. Shepard, ``Topology changing transitions in
bubbling
  geometries,'' {\em JHEP} {\bf 02} (2005) 063,
\href{http://www.arXiv.org/abs/hep-th/0502127}{{\tt
hep-th/0502127}}.
%%CITATION = HEP-TH 0502127;%%.

\bibitem{Gubser:2004xx}
S.~S. Gubser and J.~J. Heckman, ``Thermodynamics of R-charged
black holes in
  AdS(5) from effective strings,'' {\em JHEP} {\bf 11} (2004) 052,
\href{http://www.arXiv.org/abs/hep-th/0411001}{{\tt
hep-th/0411001}}.
%%CITATION = HEP-TH 0411001;%%.

\bibitem{Bak:2005ef}
D.~Bak, S.~Siwach, and H.-U. Yee, ``1/2 BPS geometries of M2 giant
gravitons,'' \href{http://www.arXiv.org/abs/hep-th/0504098}{{\tt
hep-th/0504098}}.
%%CITATION = HEP-TH 0504098;%%.

\bibitem{McGreevy:2000cw}
J.~McGreevy, L.~Susskind, and N.~Toumbas, ``Invasion of the giant
gravitons
  from anti-de Sitter space,'' {\em JHEP} {\bf 06} (2000) 008,
\href{http://www.arXiv.org/abs/hep-th/0003075}{{\tt
hep-th/0003075}}.
%%CITATION = HEP-TH 0003075;%%.

\bibitem{Hashimoto:2000zp}
A.~Hashimoto, S.~Hirano, and N.~Itzhaki, ``Large branes in AdS and
their field
  theory dual,'' {\em JHEP} {\bf 08} (2000) 051,
\href{http://www.arXiv.org/abs/hep-th/0008016}{{\tt
hep-th/0008016}}.
%%CITATION = HEP-TH 0008016;%%.

\bibitem{Balasubramanian:2005kk}
V.~Balasubramanian, V.~Jejjala, and J.~Simon, ``The library of
Babel,'' \href{http://www.arXiv.org/abs/hep-th/0505123}{{\tt
hep-th/0505123}}.
%%CITATION = HEP-TH 0505123;%%.

\bibitem{Gubser:2000nd}
S.~S. Gubser, ``Curvature singularities: The good, the bad, and
the naked,''
  {\em Adv. Theor. Math. Phys.} {\bf 4} (2002) 679--745,
\href{http://www.arXiv.org/abs/hep-th/0002160}{{\tt
hep-th/0002160}}.
%%CITATION = HEP-TH 0002160;%%.

\bibitem{Myers:1999ps}
R.~C. Myers, ``Stress tensors and Casimir energies in the AdS/CFT
  correspondence,'' {\em Phys. Rev.} {\bf D60} (1999) 046002,
\href{http://www.arXiv.org/abs/hep-th/9903203}{{\tt
hep-th/9903203}}.
%%CITATION = HEP-TH 9903203;%%.

\bibitem{Horowitz:1995ta}
G.~T. Horowitz and R.~C. Myers, ``The value of singularities,''
{\em Gen. Rel.
  Grav.} {\bf 27} (1995) 915--919,
\href{http://www.arXiv.org/abs/gr-qc/9503062}{{\tt
gr-qc/9503062}}.
%%CITATION = GR-QC 9503062;%%.

\bibitem{dobrev}
V.K. Dobrev and V.B. Petkova, ``On the group theoretical approach
to extended conformal supersymmetry: classification of
multiplets'', {\it Lett. Math. Phys.} {\bf 9}(1985), 287; V.K.
Dobrev and V.B. Petkova, ``All positive energy Unitary Irreducible
Representations of extended conformal supersymmetry'', {\it Phys.
Lett.}{\bf B162}, (1985), 127.

\bibitem{Hawking:1991nk}
S.~W. Hawking, ``The Chronology protection conjecture,'' {\em
Phys. Rev.} {\bf
  D46} (1992)
603--611.
%%CITATION = PHRVA,D46,603;%%.

\bibitem{Breckenridge:1996is}
J.~C. Breckenridge, R.~C. Myers, A.~W. Peet, and C.~Vafa,
``D-branes and
  spinning black holes,'' {\em Phys. Lett.} {\bf B391} (1997) 93--98,
\href{http://www.arXiv.org/abs/hep-th/9602065}{{\tt
hep-th/9602065}}.
%%CITATION = HEP-TH 9602065;%%.

\bibitem{Caldarelli:2004mz}
M.~M. Caldarelli, D.~Klemm, and P.~J. Silva, ``Chronology
protection in anti-de
  Sitter,''
\href{http://www.arXiv.org/abs/hep-th/0411203}{{\tt
hep-th/0411203}}.
%%CITATION = HEP-TH 0411203;%%.

\bibitem{Boni:2005sf}
M.~Boni and P.~J. Silva, ``Revisiting the D1/D5 system or bubbling
in AdS(3),'' \href{http://www.arXiv.org/abs/hep-th/0506085}{{\tt
hep-th/0506085}}.
%%CITATION = HEP-TH 0506085;%%.

\bibitem{Blau:2001ne}
M.~Blau, J.~Figueroa-O'Farrill, C.~Hull, and G.~Papadopoulos, ``A
new maximally
  supersymmetric background of IIB superstring theory,'' {\em JHEP} {\bf 01}
  (2002) 047,
\href{http://www.arXiv.org/abs/hep-th/0110242}{{\tt
hep-th/0110242}}.
%%CITATION = HEP-TH 0110242;%%.

\bibitem{Blau:2004yi}
M.~Blau, M.~Borunda, M.~O'Loughlin, and G.~Papadopoulos, ``The
universality of
  Penrose limits near space-time singularities,'' {\em JHEP} {\bf 07} (2004)
  068,
\href{http://www.arXiv.org/abs/hep-th/0403252}{{\tt
hep-th/0403252}}.
%%CITATION = HEP-TH 0403252;%%.

\bibitem{Berenstein:2002jq}
D.~Berenstein, J.~M. Maldacena, and H.~Nastase, ``Strings in flat
space and pp
  waves from N = 4 super Yang Mills,'' {\em JHEP} {\bf 04} (2002) 013,
\href{http://www.arXiv.org/abs/hep-th/0202021}{{\tt
hep-th/0202021}}.
%%CITATION = HEP-TH 0202021;%%.

\bibitem{Takayama:2005bc}
Y.~Takayama and K.~Yoshida, ``Bubbling 1/2 {BPS} geometries and
Penrose
  limits,''
\href{http://www.arXiv.org/abs/hep-th/0503057}{{\tt
hep-th/0503057}}.
%%CITATION = HEP-TH 0503057;%%.

\bibitem{Hackett-Jones:2004yi}
E.~J. Hackett-Jones and D.~J. Smith, ``Type IIB Killing spinors
and
  calibrations,'' {\em JHEP} {\bf 11} (2004) 029,
\href{http://www.arXiv.org/abs/hep-th/0405098}{{\tt
hep-th/0405098}}.
%%CITATION = HEP-TH 0405098;%%.

\bibitem{Gibbons:1999uv}
G.~W. Gibbons and C.~A.~R. Herdeiro, ``Supersymmetric rotating
black holes and
  causality violation,'' {\em Class. Quant. Grav.} {\bf 16} (1999) 3619--3652,
\href{http://www.arXiv.org/abs/hep-th/9906098}{{\tt
hep-th/9906098}}.
%%CITATION = HEP-TH 9906098;%%.

\bibitem{Herdeiro:2000ap}
C.~A.~R. Herdeiro, ``Special properties of five dimensional BPS
rotating black
  holes,'' {\em Nucl. Phys.} {\bf B582} (2000) 363--392,
\href{http://www.arXiv.org/abs/hep-th/0003063}{{\tt
hep-th/0003063}}.
%%CITATION = HEP-TH 0003063;%%.

\bibitem{Jarv:2002wu}
L.~Jarv and C.~V. Johnson, ``Rotating black holes, closed
time-like curves,
  thermodynamics, and the enhancon mechanism,'' {\em Phys. Rev.} {\bf D67}
  (2003) 066003,
\href{http://www.arXiv.org/abs/hep-th/0211097}{{\tt
hep-th/0211097}}.
%%CITATION = HEP-TH 0211097;%%.

\bibitem{Drukker:2004zm}
N.~Drukker, ``Supertube domain-walls and elimination of closed
time-like curves
  in string theory,'' {\em Phys. Rev.} {\bf D70} (2004) 084031,
\href{http://www.arXiv.org/abs/hep-th/0404239}{{\tt
hep-th/0404239}}.
%%CITATION = HEP-TH 0404239;%%.

\bibitem{Cornalba:2005je}
L.~Cornalba and M.~S. Costa, ``Unitarity in the presence of closed
timelike
  curves,''
\href{http://www.arXiv.org/abs/hep-th/0506104}{{\tt
hep-th/0506104}}.
%%CITATION = HEP-TH 0506104;%%.

\bibitem{Blau:2002mw}
M.~Blau, J.~Figueroa-O'Farrill, and G.~Papadopoulos, ``Penrose
limits,
  supergravity and brane dynamics,'' {\em Class. Quant. Grav.} {\bf 19} (2002)
  4753,
\href{http://www.arXiv.org/abs/hep-th/0202111}{{\tt
hep-th/0202111}}.
%%CITATION = HEP-TH 0202111;%%.

\bibitem{Blau:2000xg}
M.~Blau, ``Killing spinors and SYM on curved spaces,'' {\em JHEP}
{\bf 11}
  (2000) 023,
\href{http://www.arXiv.org/abs/hep-th/0005098}{{\tt
hep-th/0005098}}.
%%CITATION = HEP-TH 0005098;%%.

\bibitem{barkilling}
C.~B{\"a}r, ``Real Killing spinors and holonomy,'' {\em
Communications in
  Mathematical Physics} {\bf 154} (1993) 509--521.

\bibitem{ourconstructionforplanewave}
G.~Milanesi and M.~O'Loughlin , unpublished, 2004.

\end{thebibliography}
\end{document}